\documentclass[fleqn,usenatbib]{mnras}

\usepackage{newtxtext,newtxmath}
\usepackage[T1]{fontenc}
\usepackage{ae,aecompl}

\usepackage{graphicx}	% Including figure files
\usepackage{amsmath}	% Advanced maths commands
\usepackage{amssymb}	% Extra maths symbols
\usepackage{breqn}
\usepackage{color}
\usepackage{cases}
\usepackage{enumerate}
\usepackage{threeparttable}
\usepackage[multidot]{grffile}
\usepackage[authoryear]{natbib}
\bibpunct{(}{)}{;}{a}{}{,}

\hypersetup{draft}

\newcommand{\BlueTides }{\textsc{BlueTides} }
\newcommand{\BlueTidesns}{\textsc{BlueTides}}

\title[The host galaxies of $z=7$ quasars]{The host galaxies of $z=7$ quasars: predictions from the \BlueTides simulation}

\author[M. A. Marshall et al.]{Madeline A. Marshall$^{1,2}$\thanks{E-mail: madelinem1@student.unimelb.edu.au (MAM); tiziana@phys.cmu.edu (TDM); swyithe@unimelb.edu.au (JSBW)}, Yueying Ni$^{3}$, Tiziana Di Matteo$^{3}$\footnotemark[1], J. Stuart B. Wyithe$^{1,2}$\footnotemark[1] \newauthor
Stephen Wilkins$^{4}$, Rupert A.C. Croft$^3$, Jussi K. Kuusisto$^{4}$
\\
$^{1}$ School of Physics, University of Melbourne, Parkville, VIC 3010, Australia\\
$^{2}$ ARC Centre of Excellence for All Sky Astrophysics in 3 Dimensions (ASTRO 3D)\\
$^{3}$ McWilliams Center for Cosmology, Department of Physics, Carnegie Mellon University, Pittsburgh, PA 15213, USA \\
$^{4}$ Astronomy Centre, Department of Physics and Astronomy, University of Sussex, Brighton, BN1 9QH, UK}

\date{Accepted XXX. Received YYY; in original form ZZZ}

\pubyear{2019}

\begin{document}

\label{firstpage}
\pagerange{\pageref{firstpage}--\pageref{lastpage}}
\maketitle

% Abstract of the paper
\begin{abstract}
We examine the properties of the host galaxies
of $z=7$ quasars using the large volume, cosmological
hydrodynamical simulation \textsc{BlueTides}.
We find that the 10 most massive black holes and the 191 quasars in the simulation (with $M_{\textrm{UV,AGN}}<M_{\textrm{UV,host}}$) are hosted by massive galaxies with stellar masses $\log(M_\ast/M_\odot)=10.8\pm0.2$, and $10.2\pm0.4$, 
which have large star formation rates, of $513\substack{+1225 \\ -351}M_\odot/\rm{yr}$ and $191\substack{+288 \\ -120}M_\odot/\rm{yr}$, respectively.
The hosts of the most massive black holes and quasars in \BlueTides are generally bulge-dominated, with bulge-to-total mass ratio $B/T\simeq0.85\pm0.1$, however their morphologies are not biased relative to the overall $z=7$ galaxy sample.
We find that the hosts of the most massive black holes and quasars are compact, with half-mass radii $R_{0.5}=0.41\substack{+0.18 \\ -0.14}$ kpc and $0.40\substack{+0.11 \\ -0.09}$ kpc respectively; galaxies with similar masses and luminosities have a wider range of sizes with a larger median value, $R_{0.5}=0.71\substack{+0.28 \\ -0.25}$ kpc.
We make mock James Webb Space Telescope (JWST) images of these quasars and their host galaxies.
We find that distinguishing the host from the quasar emission will be possible but still challenging with JWST, due to the small sizes of quasar hosts. 
We find that quasar samples are biased tracers of the intrinsic black hole--stellar mass relation, following a relation that is 0.2 dex higher than that of the full galaxy sample.
Finally, we find that the most massive black holes and quasars are more likely to be found in denser environments than the typical $M_{\textrm{BH}}>10^{6.5}M_\odot$ black hole, indicating that minor mergers play at least some role in growing black holes in the early Universe.
\end{abstract}
% Select between one and six entries from the list of approved keywords.
% Don't make up new ones.

%%%%%%%%%%%%%%%%%%%%%%%%%%%%%%%%%%%%%%%%%%%%%%%%%%

%%%%%%%%%%%%%%%%% BODY OF PAPER %%%%%%%%%%%%%%%%%%
\section{Introduction}
High-redshift quasars \citep[$z\gtrsim6$,][]{Fan2000, Fan2001, Fan2003, Fan2004} are some of the most extreme systems in the Universe,
with intense accretion at or even above the Eddington limit \citep{Willott2010,DeRosa2011,DeRosa2014,Trakhtenbrot2017a} forming black holes
with masses of $10^8$--$10^9M_\odot$ \citep{Barth2003,Jiang2007,Kurk2007,DeRosa2011,DeRosa2014} in less than a billion years. These luminous
systems are invaluable probes of the early Universe, providing constraints
on black hole seed theories \citep[e.g.][]{Mortlock2011,Volonteri2012,Banados2017}, the Epoch of Reionization
\citep[e.g.][]{Fan2006a,Mortlock2011,Greig2016,Davies2018,Greig2019}, and the relation
between the growth of black holes and their host galaxies \citep[e.g.][]{Shields2006,Wang2013,Valiante2014,Schulze2014,Willott2017}.
The space density of typical Sloan Digital Sky Survey (SDSS) quasars ($M_{\textrm{UV,AGN}}<-26$) is less than 1 per $\textrm{Gpc}^{3}$ at $z\gtrsim6$ \citep{Willott2010a,Kashikawa2015,Jiang2016,Wang2019}.
Their rarity and extreme properties raise many questions such as `Are the biggest black holes found in the
rarest, most overdense regions, i.e. the biggest haloes and galaxies \citep[e.g.][]{Springel2005,Shen2007,Fanidakis2013,Ren2020}?' and `Is
this rapid growth driven by galaxy mergers, with hosts that are highly star forming,
or are their host galaxies more discy and quiet \citep[e.g.][]{Mor2012,Netzer2014,Trakhtenbrot2017}?' For further discussion see, for example, the recent reviews of \citet{Valiante2017}, \citet{Mayer2018}, and \citet{Inayoshi2020}.

Understanding the host galaxies of high-redshift quasars is essential for addressing these questions.
However, this requires detection and ideally accurate measurements of quasar host galaxies, which is extremely challenging with current telescopes \citep[see e.g.][]{Mechtley2012}.
In the rest-frame ultraviolet (UV)/optical, which traces the emission from the accretion disc and the
host stellar component, the quasars often outshine their hosts, entirely concealing the host galaxy emission \citep{Mechtley2012}. The detection of $z\gtrsim6$ quasar
host galaxies has indeed eluded the Hubble Space Telescope (HST).
The only current detections of high-redshift quasar hosts are instead in the rest-frame far-infrared, 
observed in the sub-mm \citep[e.g.][]{Bertoldi2003,Walter2003,Walter2004,Riechers2007,Wang2010,Wang2011,Venemans2019}, which traces cold dust in the host galaxy.

Observations in sub-mm and mm wavelengths with the Atacama Large Millimeter Array (ALMA) and the IRAM Plateau de Bure Interferometer (PdBI), for example, imply a diverse population of quasar hosts, with
inferred dynamical masses of $10^{10}$--$10^{11}M_\odot$ \citep{Walter2009,Wang2013,Venemans2015,Venemans2017a,Willott2017,Trakhtenbrot2017,Izumi2018,Izumi2019,Pensabene2020,Nguyen2020}, 
dust masses of $10^7$--$10^9M_\odot$ \citep{Venemans2015,Izumi2018,Nguyen2020},
sizes of 1--5 kpc \citep{Wang2013,Venemans2015,Willott2017,Izumi2019}, and a wide range of star-formation rates (SFRs) of $10$--$3000 M_\odot/$yr \citep{Venemans2015,Venemans2017a,Willott2017,Trakhtenbrot2017,Izumi2018,Izumi2019,Shao2019,Nguyen2020}.
The hosts are found in a variety of dynamical states, with some
having nearby companions which may suggest a merger system \citep[e.g][]{Trakhtenbrot2017}, while some show signatures of a rotating disc \citep[e.g][]{Willott2017,Trakhtenbrot2017}, or even no ordered motion \citep{Venemans2017}. 
However, since cold dust
may not trace the stellar distribution, there may be significant biases in stellar properties inferred through these observations \citep[e.g][]{Narayanan2009,Valiante2014,Lupi2019}.

Upcoming facilities will provide the next frontier for understanding high-redshift
quasars and their host galaxies. Infrared surveys with Euclid \citep{Amiaux2012} and the Nancy Grace Roman Space Telescope \citep[RST, formerly the Wide Field Infrared Survey Telescope or WFIRST;][]{Spergel2015} will significantly increase the known sample of $z\gtrsim6$ quasars. 
The improved resolution of the James Webb Space Telescope \citep[JWST;][]{Gardner2006} will allow the first
detections of the stellar component of their host galaxies, which will be invaluable for
accurately determining the properties of quasar hosts. Making detailed theoretical
predictions for the results of these groundbreaking instruments is thus a current priority.

Due to the rarity of high-redshift quasars, comprehensive theoretical predictions
require high-resolution simulations with large computational volumes.
Cosmological hydrodynamical simulations such as Massive Black \citep{Matteo2012}, with a volume of
(0.76 Gpc$)^3$, and \BlueTides \citep{Feng2015}, with a volume of (0.57 Gpc$)^3$, have pioneered
this area. These simulations have been used to investigate the rapid growth of black holes \citep{Matteo2012,DeGraf2012a,Feng2014,Matteo2017} and their relationship
to their host galaxies \citep{Khandai2012,DeGraf2015,Huang2018}, and make predictions for the highest-redshift quasars that are observed \citep{DeGraf2012,Tenneti2018,Ni2018}.

Previous \BlueTides analyses were performed with the phase I simulation, which reached a minimum redshift of $z=8.0$, and \textsc{BlueTides-II}, the second phase of the simulation which had been run to $z=7.5$ when last analysed \citep{Tenneti2018}.
In this paper we use the \textsc{BlueTides-II} simulation extended further to $z=7.0$ to make predictions for the properties
of quasar host galaxies. At $z=7.5$, there is one quasar analogue in the \BlueTides simulation, as studied by \citet{Tenneti2018}. Extending the simulation from $z=7.5$ to 7.0, a period of only 58 Myr, results in a
considerable increase in the number of observable quasar analogues to the order of 100, since this is such an intense growth period for black holes in the Universe. This statistical
sample allows us to make predictions for the broader quasar population, and not
just for individual, extreme systems as was possible previously. 

The paper is outlined as follows. In Section \ref{sec:Simulation} we describe the
simulation and the post-processing used to obtain mock spectra of the
quasars and their host galaxies. In Section \ref{sec:intrinsic} we consider the intrinsic galaxy properties of the hosts of black holes and quasars. 
We consider observable properties in Section \ref{sec:Observations}, making spectra and mock JWST images.
In Section \ref{sec:Relation} we examine the black hole--stellar mass relation,
showing how observations of these quasars will lead to a biased measurement. 
We explore the environments of quasars in Section \ref{sec:Mergers}, before concluding in 
Section \ref{sec:Conclusions}.
The cosmological parameters used throughout are from the nine-year Wilkinson Microwave Anisotropy Probe \citep[WMAP;][]{Hinshaw2013}: $\Omega_M=0.2814$, $\Omega_\Lambda=0.7186$, $\Omega_b=0.0464$, $\sigma_8=0.820$, $\eta_s=0.971$ and $h=0.697$.

\section{Simulation}
\label{sec:Simulation}
\subsection{\BlueTides}

The \BlueTides simulation\footnote{http://BlueTides-project.org/} \citep{Feng2015} is a cosmological hydrodynamical simulation, which uses the Pressure Entropy Smoothed Particle Hydrodynamics (SPH) code MP-Gadget to model the evolution of $2\times 7040^{3}$ particles in a cosmological box of volume $(400/h ~\rm{cMpc})^3$.
The mass resolution of the simulation is $1.2 \times 10^7/h~ M_{\odot}$ for dark matter particles and $2.4 \times 10^6/h~ M_{\odot}$ for gas particles (in the initial condition). Star particles are converted from gas particles with sufficient star formation rates, and each have a stellar mass of $6\times10^{5}/h~ M_\odot$. The gravitational softening length of $\epsilon_{\rm grav} = 1.5/h \textrm{~ckpc}$ is the effective spatial resolution.
From the initial conditions at $z=99$, \BlueTides evolved the box to $z=8$ in phase I \citep{Feng2015}.
Phase II of the simulation continued the evolution of the box from $z=8$ to lower redshifts, with the first results from this phase given in \citet{Tenneti2018}. Here we focus on the lowest redshift currently reached by phase II, $z=7.0$.
From the simulation, we consider the 108,000 most massive halos, with masses $M_{\textrm{vir}}>10^{10.8} M_\odot$, which contain galaxies with $M_\ast>10^{5.9} M_\odot$ and black holes with $M_{\textrm{BH}}>10^{5.8}M_\odot$ (the seed mass).

\BlueTides implements a variety of sub-grid physics to model galaxy and black hole formation and their feedback processes.
Here we briefly list some of its basic features, and refer the reader to the original paper~\citep{Feng2015} for more detailed descriptions. 
In the \BlueTides simulation, gas cooling is performed through both primordial radiative cooling~\citep{Katz} and metal line cooling~\citep{Vogelsberger2014}. 
Star formation is based on the multi-phase star formation model originally from \citet{SH03} with modifications following~\citet{Vogelsberger2013}.
We also implement the formation of molecular hydrogen and model its effects on star formation using the prescription from \citet{Krumholtz}, where we self-consistently estimate the fraction of molecular hydrogen gas from the baryon column density, which in turn couples the density gradient into the star formation rate.
For stellar feedback, we apply a type-II supernova wind feedback model from \citet{Okamoto}, assuming wind speeds proportional to the local one-dimensional dark matter velocity dispersion. 
The large volume of \BlueTides also allows the inclusion of a model of `patchy reionization' ~\citep{Battaglia}, yielding a mean reionization redshift $z\simeq10$, and incorporating the UV background estimated by \citet{fg09}. 

The black hole sub-grid model associated with black hole growth and active galactic nuclei (AGN) feedback is the same as that in the \textsc{MassiveBlack I \& II} simulations, originally developed in \citet{SDH2005} and \citet{DSH2005}, with modifications consistent with \textsc{Illustris}; see \citet{DeGraf2012a} and \citet{DeGraf2015} for full details.
Black holes are seeded with a mass of $M_{\textrm{BH,seed}} = 5 \times 10^5/h~ M_{\odot}$ in dark matter haloes above a threshold mass of $M_{\textrm{Halo}} = 5 \times 10^{10}/h~ M_{\odot}$. The simulation makes no direct assumption of the black hole formation mechanism, although this mass is most consistent with seed masses predicted by direct collapse scenarios \citep[e.g.][]{Begelman2006,Shang2010,Volonteri2010,Latif2013}.
Black holes grow by merging with other black holes, and via gas accretion at the Bondi-Hoyle accretion rate \citep{Hoyle1939,Bondi1944,Bondi1952}, $\dot{M_{\textrm{BH}}}=\alpha 4\pi G^2 M_{\textrm{BH}}^2 \rho_{\textrm{BH}} (c_s^2+v^2)^{-3/2}$, where $\rho_{\textrm{BH}}$ is the local gas density, $c_s$ is the local sound speed, $v$ is the velocity of the black hole relative to the surrounding gas, and $\alpha$ is a dimensionless parameter.
Mildly super-Eddington accretion is permitted, with the accretion rate limited to two times the Eddington limit. 
In this sub-grid model, the black hole mass grows smoothly at this accretion rate.
In order to account for the discrete nature of gas particles, once a black hole has grown by an amount equivalent to the mass of a gas particle, a gas particle is removed and its mass transferred to the black hole's `dynamical mass'. This discrete dynamical mass allows the particle dynamics to be calculated correctly within the simulation; however, the continuous black hole mass is always used in any analyses.
Finally, we assume that black holes radiate with bolometric luminosity $L_{\textrm{AGN}}=\eta \dot{M}_{\textrm{BH}} c^2$, with a radiative efficiency $\eta$ of $0.1$. 

Throughout this work, we generally consider only black holes with $M_{\textrm{BH}}>10^{6.5}M_\odot$, in order to minimize any possible influence of the seeding prescription on our analysis. We also consider only the $z=7.0$ snapshot. This is in contrast to \citet{Tenneti2018}, which explored a range of snapshots around $z\simeq7.5$ in order to find the brightest quasar in the simulation, as quasar luminosity varies significantly due to the time-variability of black hole accretion.
Using only one snapshot is more representative of an observational sample, in which galaxies are observed at a random phase in their growth history, and not, for example, only when their black hole is at its peak luminosity.

To extract the properties of galaxies from the simulation, we run a friends-of-friends (FOF) algorithm \citep{Davis1985}. The galaxy properties, such as the star formation density, stellar mass function and UV luminosity function, have been shown to match current observational constraints at $z=8$, 9 and 10 \citep{Feng2015,Waters2016,Wilkins2017}.

To determine the stellar mass of the galaxies from the total stellar mass contained in their host dark matter haloes, we calculate the galaxy $R_{200}$, the radius containing 200 times the critical stellar mass density (the critical density of the Universe multiplied by the baryon fraction and star formation efficiency of the simulation). 
We define the stellar mass of a galaxy as the mass contained within $R_{200}$.
This generally includes the inner dense core of the galaxy and also the more diffuse outer regions, ensuring that we include the majority of particles truly associated with each galaxy.
For determining the sizes of galaxies, we calculate the half-mass radius inside this $R_{200}$, $R_{0.5}$.
%This restricts the calculation to the local density peak around the galaxy, ensuring that the mass from neighbouring galaxies is excluded.  $3 R_{0.5}$ \citep[as in, e.g.][]{Tacchella2019}.

We determine the morphology of the galaxy by its bulge-to-total ratio, calculated using the bulge-to-disc decomposition method of \citet{Scannapieco2009}. We first construct a circularity parameter $\epsilon = j_z/j_{\rm circ}(r)$ for each star particle in the galaxy within $R_{200}$, where $j_z$ is the projection of the specific angular momentum of the star particle in the direction of the total angular momentum of the galaxy, and $j_{\rm circ}(r)$ is the angular momentum expected for a circular orbit at the radius $r$: $j_{\rm circ} = r v_{\rm circ}(r) = \sqrt{GM(<r)r}$. We identify star particles with $\epsilon > 0.7$ as disc stars, and define the bulge-to-total ratio as $B/T = 1 - f_{\epsilon > 0.7}$, where $f_{\epsilon > 0.7}$ is the fraction of disc stars in the galaxy.

\subsection{Mock spectra}

\subsubsection{Galaxy SEDs}
To determine the spectral energy distribution (SED) of a galaxy, we assign a SED from a simple stellar population (SSP) to each star particle within $R_{200}$, based on its mass, age and metallicity. We do this using the Binary Population and Spectral Population Synthesis model \citep[BPASS, version 2.2;][]{Stanway2018}, assuming a modified Salpeter initial mass function with a high-mass cut-off of $300M_\odot$. The SED of the galaxy is taken as the sum of the SEDs of each of its star particles. To determine the relative contribution of the stellar and nebular emission, we assume an escape fraction of 0.9.

\subsubsection{Quasar spectra}
To assign spectra to each of our quasars, we use the CLOUDY spectral synthesis code \citep{Ferland2017}, as in \citet{Tenneti2018}.

The continuum is given by
\begin{equation}
f_\nu = \nu^{\alpha_{\textrm{UV}}} \exp \left(\frac{-h\nu}{kT_{\textrm{BB}}}\right) \exp \left(\frac{-kT_{\textrm{IR}}}{h\nu}\right) +a \nu^{\alpha_{\textrm{X}}}
\end{equation}
where $\alpha_{\textrm{UV}}=-0.5$, $\alpha_{\textrm{X}}=-1$, $kT_{\textrm{IR}}=0.01$Ryd, and $T_{\textrm{BB}}$ is the temperature of the accretion disc, which is determined by the black hole mass and its accretion rate
\begin{equation}
T_{\textrm{BB}}=\left( \frac{3c^6}{8\pi6^3 \sigma_{\textrm{SB}}G^2}   \frac{\dot{M}_{\textrm{BH}}}{M_{\textrm{BH}}^2}  \right)^{1/4} = 2.24\times 10^9 \left( \frac{\dot{M}_{\textrm{BH}}}{M_\odot/\textrm{yr}} \right)^{1/4} \left( \frac{M_{\textrm{BH}}}{M_\odot} \right)^{-1/2} \textrm{K} .
\end{equation}
The normalization of the continuum is set by the bolometric luminosity of the quasar.

The emission lines are calculated with CLOUDY assuming a hydrogen density of $10^{10} \textrm{cm}^{-3}$ at the face of the cloud, which has inner radius $10^{18}$cm, and a total hydrogen column density of $10^{22} \textrm{cm}^{-2}$.

We also implement Lyman-forest extinction on the redshifted spectra \citep{Madau1995,synphot} for both the quasars and the host galaxies.

\subsubsection{Dust attenuation and extinction}
\label{sec:dust}
As in \citet{Wilkins2017}, we model the dust attenuation of galaxies by relating the density of metals along a line of sight to the UV-band dust optical depth $\tau_{\rm UV}$. 
For each star particle in the galaxy, we calculate $\tau_{\rm UV,\ast}$ as
\begin{equation}
\label{eq:dust_extinction}
 \tau_{{\rm UV,}\ast}= -\kappa \Sigma(x, y, z) \left(\frac{\lambda}{5500\text{\normalfont\AA}}\right)^{\gamma}.
\end{equation}
where $\Sigma(x, y, z) = \int_{z'=0}^{z} \rho_{\rm metal}(x,y,z')dz'$ is the metal surface density at the position of the star particle, along the z-direction line of sight, and $\kappa$ and $\gamma$ are free parameters. 
Here we use $\kappa=10^{4.6}$ and $\gamma=-1.0$, which are calibrated against the observed galaxy UV luminosity function at redshift $z=7$.
The total dust-attenuated galaxy luminosity is the sum of the extincted luminosities of each individual star particle.

We apply the same technique to determine the dust attenuation of the AGN, with the dust optical depth calculated using the metal column density integrated along a line of sight to the quasar:
\begin{equation}
\label{equation:tau_dust}
    \tau_{\rm UV, AGN} = \kappa  \int_{\rm ray} \rho_{\rm metal}(l) dl \left(\frac{\lambda}{5500\text{\normalfont\AA}}\right)^{\gamma},
\end{equation}
with the same values of $\kappa$ and $\gamma$ \citep[see also][]{Ni2019}.
Dust attenuation of the AGN mainly traces the regions of high gas density near the centre of the galaxy, with the gas metallicity $Z$ only modulating the dust extinction at a sub-dominant level \citep[see Figure 14 in][for illustration.]{Ni2019}.  
Because of the angular variation in the density field surrounding the central black hole, the dust extinction for AGN is sensitive to the choice of line of sight, unlike for galaxies, whose dust attenuation is accumulated over the extended source.
For each quasar we therefore calculate $\tau_{\rm UV, AGN}$ along approximately 1000 lines of sight. See \citet{Ni2019} for full details.
%(That is not the case for galaxy dust extinction though: Since galaxy is extended source, dust attenuation is an accumulative effect over the parallel lines of sights through the galaxy and therefore would not vary a lot with respect to different viewing angles.)

\section{Properties of black hole host galaxies}
\label{sec:intrinsic}
\subsection{Sample selection}
\label{sec:Sample}

\begin{figure*}
\begin{center}
\includegraphics[scale=0.9]{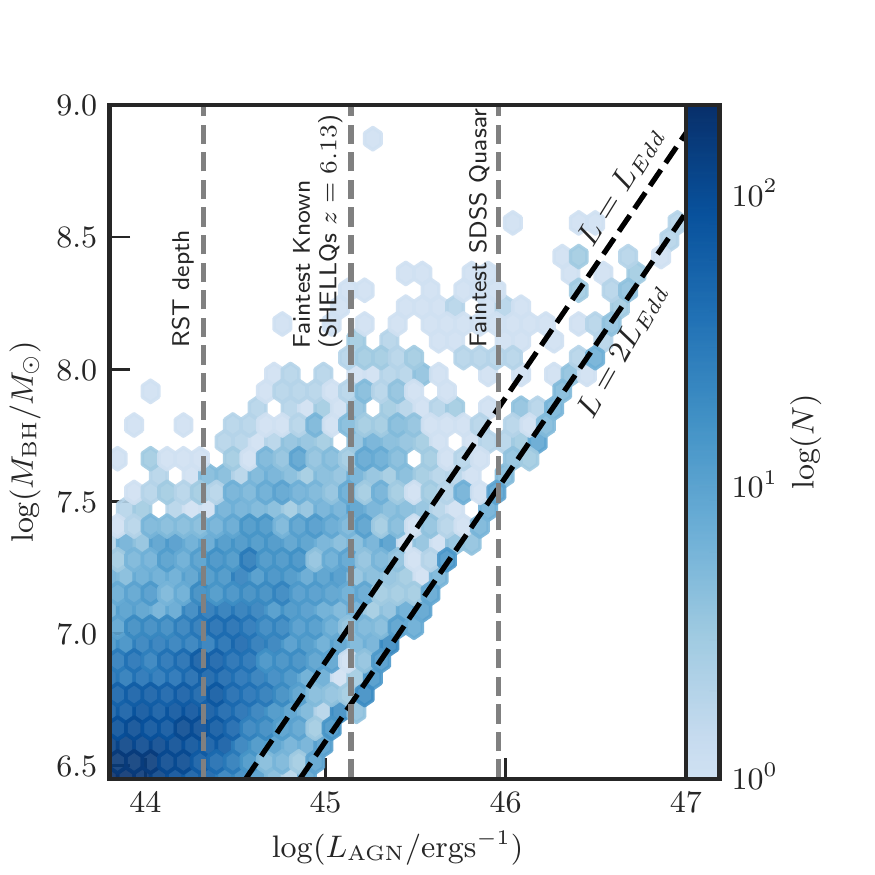}
\includegraphics[scale=0.53]{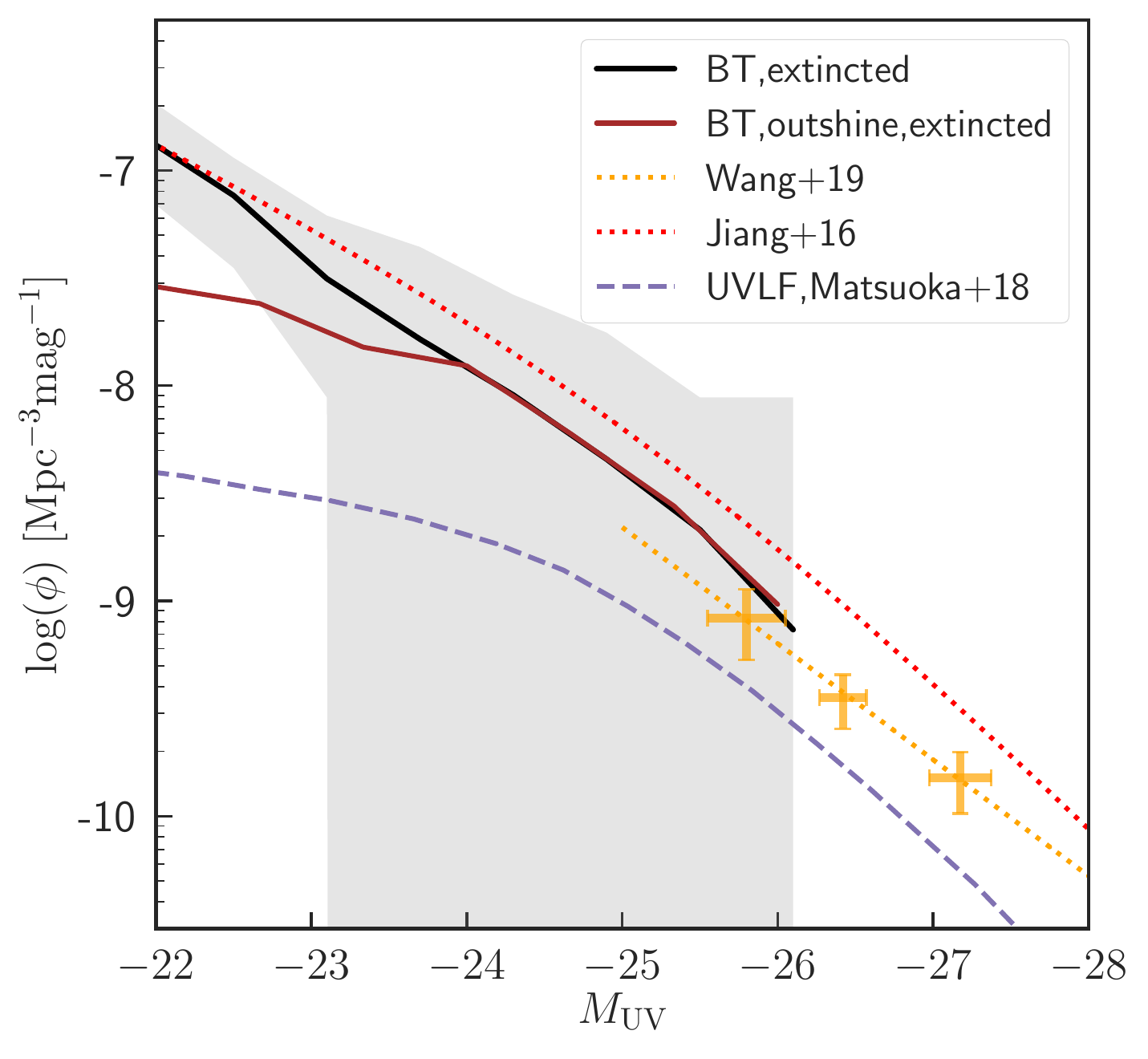}
\caption{\textit{Left:} The distribution of black hole masses and AGN bolometric luminosities for \BlueTides galaxies at $z=7$ (blue density plot). The Eddington limit is shown for reference (dashed black line), as well as twice the Eddington limit, which is the upper limit of the black hole accretion rate set in the simulation. The luminosity of the faintest SDSS quasar, the faintest currently-known high-redshift quasar, and the RST detection limit are shown for reference (grey dashed lines).
\textit{Right}: The UV-band luminosity function of AGN at $z=7$. The black solid line is the dust-extincted quasar luminosity function from \textsc{BlueTides}. The brown solid line is the luminosity function including only the AGN that outshine their host galaxies.
The grey shaded area gives the error estimate by considering the dust extinction through all lines of sight of the AGN population. 
The orange solid symbols with error bars show the measured binned quasar luminosity function from \citet{Wang2019} at $z \simeq 6.7$.
The red dotted line is the $z \simeq 6$ fitted quasar luminosity function measured by \citet{Jiang2016}. 
The purple dashed line gives the luminosity function from \citet{Matsuoka2018b}, based on the population of $5.7<z<6.5$ quasars and extrapolated to $z=7$.
}
\label{BHmassluminosity}
\end{center}
\end{figure*}

The black hole population in \BlueTides at $z=7.0$ is presented in Figure \ref{BHmassluminosity}, which shows the distributions of their masses and AGN luminosities, as well as the AGN UV luminosity function. 
By considering the UV-band dust extinction as described in Section \ref{sec:dust} and \citet{Ni2019}, \BlueTides produces a quasar luminosity function that is in good agreement with the high-redshift observations of \citet{Jiang2016} and \citet{Wang2019} ---\BlueTides predicts the expected number density of high-redshift quasars. 
The most massive black holes in \BlueTides at $z=7$ have masses of $M_{\textrm{BH}}\simeq10^{8.5}M_\odot$, and the most luminous AGN have intrinsic bolometric luminosities $L_{\textrm{bol}}\simeq10^{47} \textrm{erg s}^{-1} \simeq 10^{13}L_\odot$, equivalent to those of currently observed high-redshift quasars.

\begin{figure}
\begin{center}
\includegraphics[scale=0.9]{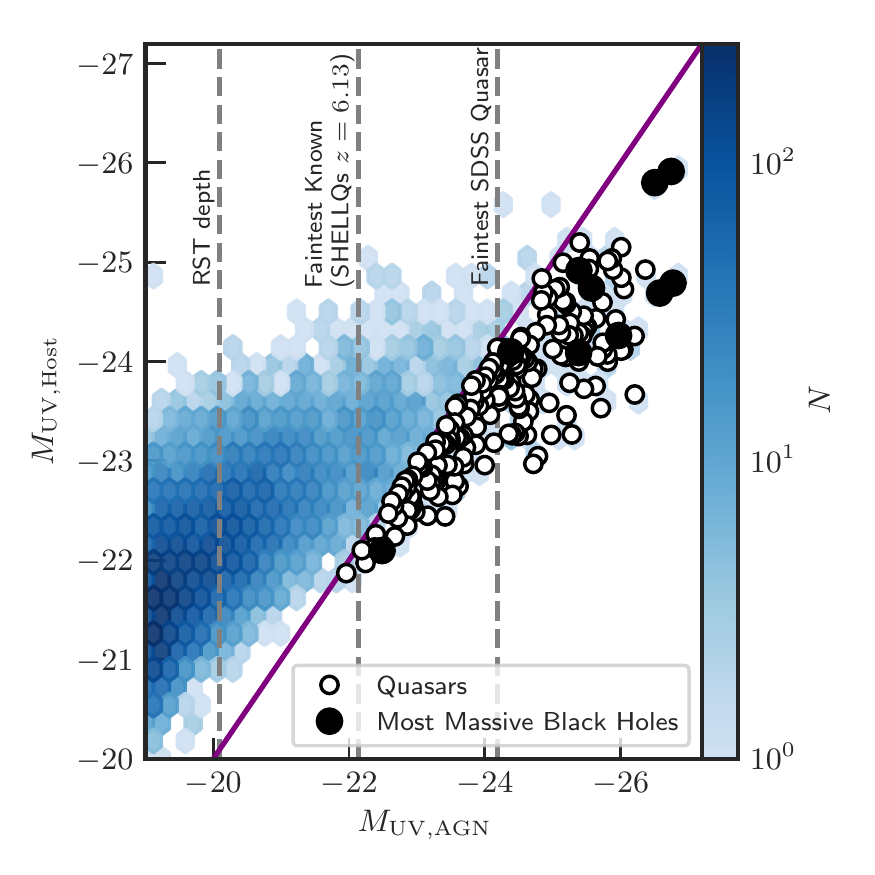}
\caption{The distribution of host and AGN intrinsic UV absolute magnitudes for \BlueTides galaxies at $z=7$ (blue density plot). 
We classify quasars (white circles) as those with $M_{\textrm{UV,AGN}}<M_{\textrm{UV,Host}}$, since the AGN outshines the host galaxy. The 10 most massive black holes (black circles) are also shown. The luminosity of the faintest SDSS quasar, the faintest currently-known high-redshift quasar, and the RST detection limit are shown for reference (grey dashed lines).}
\label{Muv_gal_agn}
\end{center}
\end{figure}

From this black hole population we select three samples; the most massive black holes, `quasars', and `hidden quasars'.
\\
\textbf{Massive black hole sample:}
We consider the ten most massive black holes at $z=7$, which have masses $M_{\textrm{BH}}=10^{8.44}-10^{8.89}M_\odot$, to be our `massive black hole' sample. 
\\
\textbf{Quasar sample:}
To select `quasars' from the simulation, we consider the galaxy and AGN UV-band absolute magnitudes, as shown in Figure \ref{Muv_gal_agn}. 
We make the simple assumption that every bright ($L_{\textrm{AGN}}>10^{44} \textrm{erg s}^{-1}$) black hole with $M_{\textrm{UV,AGN}}<M_{\textrm{UV,Host}}$ would be classified as a quasar, since the AGN outshines the host galaxy. 
This results in a sample of 205 \BlueTides quasars, which is only 2.1 per cent of the black holes with $M_{\textrm{BH}}>10^{6.5}M_\odot$, and 2.6 per cent of black holes with $L_{\textrm{AGN}}>10^{44} \textrm{erg s}^{-1}$ (see Figure \ref{Muv_gal_agn}).
We note that the assumption that $M_{\textrm{UV,AGN}}<M_{\textrm{UV,Host}}$ for a galaxy to be classified as a quasar is not an accurate representation of the true observational quasar selection techniques, and may underestimate the number of galaxies in our sample that would be observed as quasars. 
\\
\textbf{Hidden quasar sample:}
Within our simulation, 70.5 per cent of black holes brighter than the faintest currently-known high-redshift quasar, $m_{\textrm{UV}}=24.85$ \citep[$L_{\textrm{AGN}}>10^{45.1} \textrm{erg s}^{-1}$ at $z=7$;][]{Matsuoka2018}, have host luminosities that outshine the AGN. 
These 488 black holes are experiencing significant black hole growth, with high AGN luminosities, but are simply `hidden' by their luminous host galaxies. 
We consider all black holes with intrinsic luminosities $L_{\textrm{AGN}}>10^{45.1} \textrm{erg s}^{-1}$ and with $M_{\textrm{UV,AGN}}>M_{\textrm{UV,Host}}$ as `hidden' quasars, i.e. those outshined by their host galaxy.

\subsection{Galaxy properties}
\label{sec:Properties}
We now investigate the properties of the hosts of the most massive black holes and quasars in \BlueTides at $z=7$. 

\begin{figure}
\begin{center}
\includegraphics[scale=0.9]{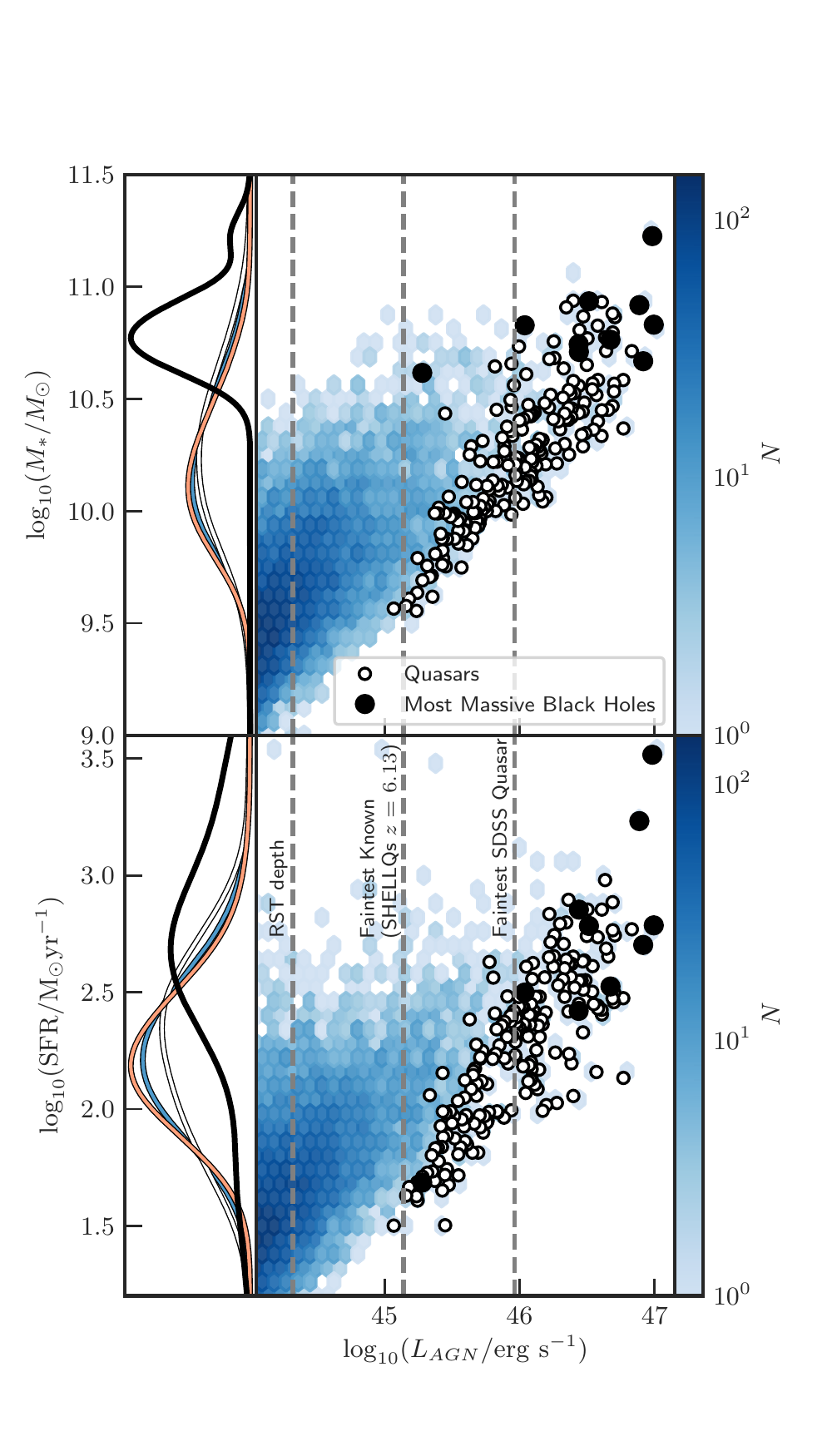}
\caption{The relations between AGN luminosity and stellar mass (upper right panel) and star formation rate (lower right panel). The blue density plot shows the distribution for all \BlueTides galaxies, with the most massive black holes and quasars also plotted (see legend).
The left panels show the distributions of the host properties for the most massive black holes (black line), quasars (white line), hidden quasars (salmon line), and for all black holes with $L_{\textrm{AGN}}>10^{45.1} \textrm{erg s}^{-1}$ (blue line). The luminosity of the faintest SDSS quasar, the faintest currently-known high-redshift quasar ($L_{\textrm{AGN}}=10^{45.1}\textrm{erg s}^{-1}$), and the RST detection limit are shown in the right panels for reference (grey dashed lines).}
\label{fig:stellarmass_sfr}
\end{center}
\end{figure}

\begin{figure}
\begin{center}
\includegraphics[scale=0.9]{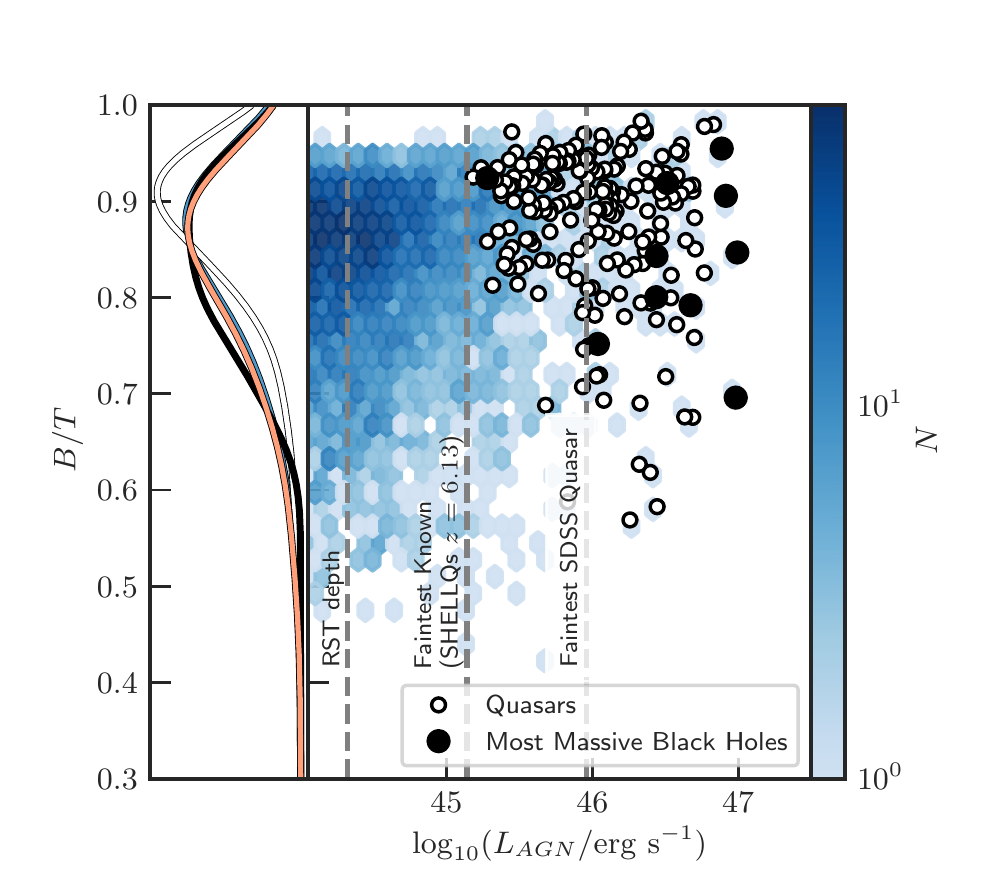}
\caption{The relation between AGN luminosity and the bulge-to-total mass ratio ($B/T$). The blue density plot shows the distribution for all \BlueTides galaxies, with the most massive black holes and quasars also plotted (see legend). The left panel shows the distribution of $B/T$ for the most massive black holes (black line), quasars (white line), hidden quasars (salmon line), and all black holes brighter than $L_{\textrm{AGN}}>10^{45.1} \textrm{erg s}^{-1}$ (blue line). The luminosity of the faintest SDSS quasar, the faintest currently-known high-redshift quasar ($L_{\textrm{AGN}}=10^{45.1}\textrm{erg s}^{-1}$), and the RST detection limit are shown in the right panel for reference (grey dashed lines).}
\label{fig:BulgeToTotal}
\end{center}
\end{figure}

\begin{figure*}
\begin{center}
\includegraphics[scale=0.95]{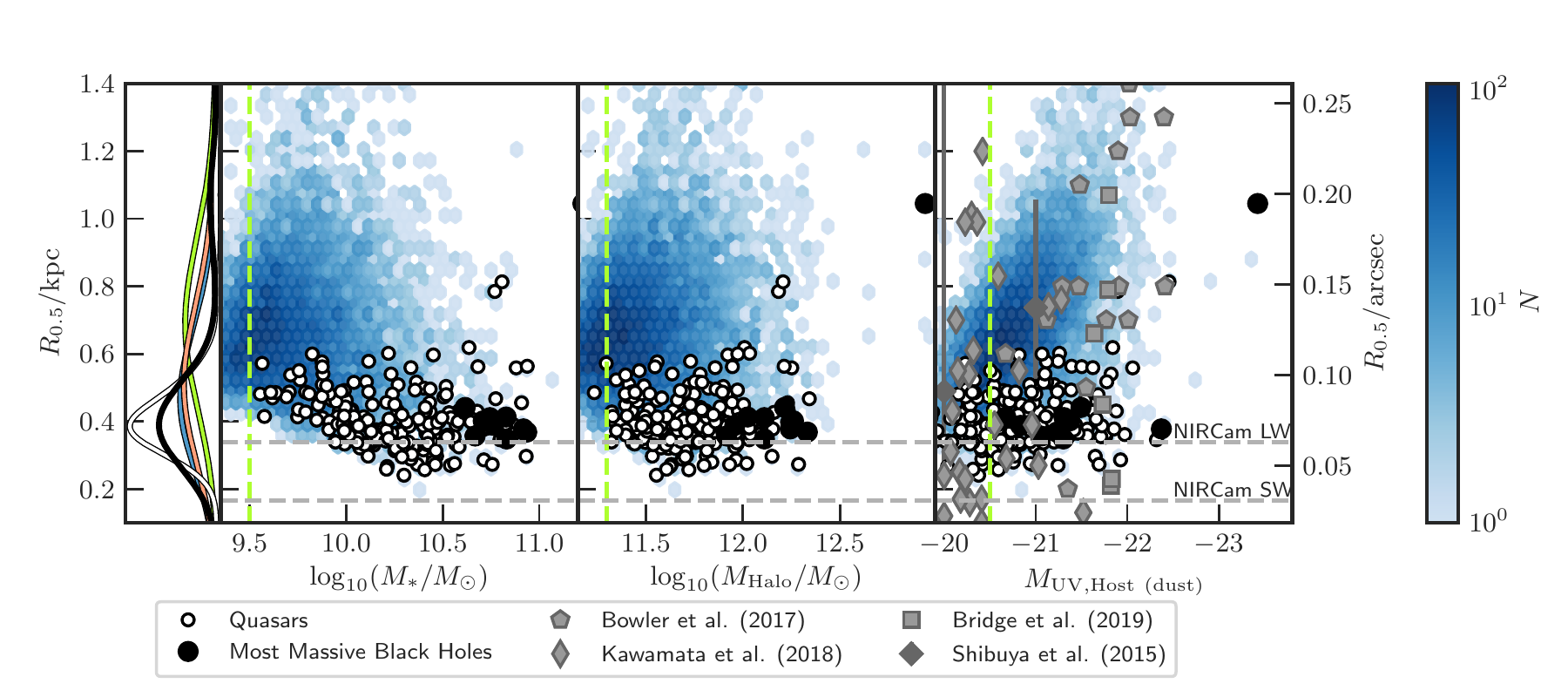}
\caption{The relation between half-mass radius and stellar mass (left), halo mass (centre), and dust-attenuated galaxy UV magnitude (right). The blue density plots show the distribution for all \BlueTides galaxies, with the most massive black holes and quasars also plotted (see legend). Also shown in the right panel are a range of observations of individual $z\simeq6-7$ Lyman-break galaxies \citep{Bowler2016,Kawamata2018,Bridge2019}, and the size--luminosity relation for Lyman-break galaxies at $z=7$ derived by \citet{Shibuya2015}. Horizontal grey dashed lines show the pixel scales of the JWST NIRCam short-wavelength (SW; 0.6--2.3 $\mu$m) and long-wavelength (LW; 2.4--5.0 $\mu$m) detectors, of $0\farcs031$ and $0\farcs063$ respectively, for reference.
The left-most panel shows the distributions of half-mass radius for the most massive black holes (black line), quasars (white line), hidden quasars (salmon line), all black holes brighter than $L_{\textrm{AGN}}>10^{45.1} \textrm{erg s}^{-1}$ (blue line) and the total sample of galaxies with $M_\ast>10^{9.5} M_\odot$, $M_{\textrm{Halo}}>10^{11.3} M_\odot$ and $M_{\textrm{UV,Host~(dust)}}<-20.5$ (green line). These limits are shown in the corresponding panels for reference (green dashed lines). }
\label{fig:Size}
\end{center}
\end{figure*}

In Figure \ref{fig:stellarmass_sfr} we show the relation between AGN luminosity and both stellar mass and star formation rate at $z=7$.
The most massive black holes are in massive galaxies with stellar masses $\log(M_\ast/M_\odot)=10.80\substack{+0.20 \\ -0.16}$, 
which have large star formation rates, $513\substack{+1225 \\ -351}M_\odot/\rm{yr}$.\footnote{Errors presented in this manner correspond to the 16th and 84th percentiles of the distributions, relative to the median value} 
We find that the quasar hosts also have large but lower stellar masses of $\log(M_\ast/M_\odot)=10.25\substack{+0.40 \\ -0.37}$, 
and lower star formation rates of $191\substack{+288 \\ -120}M_\odot/\rm{yr}$.

These star formation rates are broadly consistent with those observed in the hosts of luminous high-redshift quasars with the PdBI of $\simeq 1700 M_\odot/\rm{yr}$ \citep{Walter2009}, and with ALMA: 
100--1600 $M_\odot/\rm{yr}$ \citep{Venemans2015},
200--3500 $M_\odot/\rm{yr}$ \citep{Trakhtenbrot2017},
30--3000 $M_\odot/\rm{yr}$ \citep{Decarli2018},
50--2700 $M_\odot/\rm{yr}$ \citep{Venemans2018},
900--3200 $M_\odot/\rm{yr}$ \citep{Nguyen2020},
and $\simeq 2500 M_\odot/\rm{yr}$ \citep{Shao2019}. However, 
the simulation does not contain quasar hosts with extreme star formation rates of $\gtrsim1000M_\odot/\rm{yr}$, as are observed. 
This is most likely because by $z=7$~ \BlueTides has not yet produced a population of extremely luminous quasars, which are those generally found in such extreme hosts.
Note, however, that star-formation rates derived from far-infrared observations can have uncertainties of a factor of $\sim2$--3 \citep[e.g.][]{Venemans2018}. A comparison of \BlueTides at lower redshift with more precise star formation rates measured in the rest-frame UV using JWST, for example, would allow for a deeper understanding of quasar host star formation rates.

From Figure \ref{fig:stellarmass_sfr} we see that, on average, lower luminosity quasars have less extreme host galaxies, with lower masses and star formation rates.
The hosts of lower luminosity high-redshift quasars are indeed observed to have lower star formation rates:
$\lesssim10 M_\odot/\rm{yr}$ \citep{Willott2017}, 100--500 $M_\odot/\rm{yr}$ \citep{Trakhtenbrot2017}, 23--40 $M_\odot/\rm{yr}$ \citep{Izumi2018}, and 200--500 $M_\odot/\rm{yr}$ \citep{Nguyen2020}. While our $z=7$ predictions do not extend to such low star formation rates, we expect that by $z\simeq6$ there could be more scatter in the relation, alongside more `quenched' quasars, where feedback has significantly reduced the star formation in the host galaxy.

Figure \ref{fig:stellarmass_sfr} also shows the one-dimensional distributions of stellar mass and star formation rate for these samples, alongside all black holes brighter than $L_{\textrm{AGN}}>10^{45.1} \textrm{erg s}^{-1}$, and `hidden' quasars. This shows that the most massive black holes live in more massive galaxies, with higher star formation rates, than the total sample of bright black holes ($L_{\textrm{AGN}}>10^{45.1} \textrm{erg s}^{-1}$), which have $\log(M_\ast/M_\odot)=10.18\substack{+0.37 \\ -0.32}$ and star formation rates of $170\substack{+219 \\ -90}M_\odot/\rm{yr}$. 
The hidden quasars are hosted by galaxies with $\log(M_\ast/M_\odot)=10.16\substack{+0.34 \\ -0.30}$ and star formation rates of $166\substack{+189 \\ -83}M_\odot/\rm{yr}$.

At a fixed AGN luminosity, the quasars are hosted by less massive galaxies with lower star formation rates than the hidden quasars. This is expected due to the $M_{\textrm{UV,AGN}}<M_{\textrm{UV,Host}}$ selection: the quasar sample contains galaxies with lower $M_{\textrm{UV,Host}}$ for fixed $M_{\textrm{UV,AGN}}$, which is produced by having a lower star formation rate. Galaxies with higher star formation rates have higher luminosities which outshine their quasar, resulting in `hidden' quasars of the same quasar luminosity. As stellar mass is an integrated quantity, the selection effect is weakened slightly. 

In Figure \ref{fig:BulgeToTotal} we show the relation between AGN luminosity and the ratio of stellar mass contained in a galaxy's bulge to its total stellar mass ($B/T$). The hosts of the most massive black holes and quasars all show bulge-dominated morphologies, although there is a large tail to lower $B/T$, with $B/T=0.85\substack{+0.09 \\ -0.10}$, and $0.89\substack{+0.07 \\ -0.10}$ for the two samples respectively.
Their morphologies have a similar distribution to that of the total sample of bright black holes ($L_{\textrm{AGN}}>10^{45.1} \textrm{erg s}^{-1}$), with $B/T=0.85\substack{+0.09 \\ -0.12}$, and hidden quasars, with $B/T=0.84\substack{+0.10 \\ -0.14}$. 
The hosts of the most massive black holes and quasars in \BlueTides are generally bulge-dominated, but are not biased in morphology relative to the overall galaxy sample at $z=7$.

\citet{Lupi2019} performed a high-resolution cosmological zoom-in simulation of a halo containing a black hole with mass $M_{\textrm{BH}}=10^{8.9}M_\odot$ at $z=7$, similar to that of the most massive black hole in \textsc{BlueTides}. They found its host galaxy to have a mass of $M_\ast\simeq10^{11}M_\odot$ and a large star formation rate of $\sim 10^{2.5}M_\odot/\rm{yr}$ at $z=7$, equivalent to the most massive and star forming galaxies in \textsc{BlueTides}. Their quasar host is less bulge-dominated than those in our quasar sample, with a bulge-to-total mass ratio of $B/T\simeq0.45$. This is potentially due to the increased resolution of their simulation, which has the ability to better resolve the disc structure.
The results of \citet{Lupi2019} are therefore reasonably consistent with the \BlueTides simulation, given their sample of only one quasar host.

Figure \ref{fig:Size} shows the relation between half-mass radius $R_{0.5}$ and stellar mass, halo mass and host UV magnitude. For comparison, we also show a range of observations of $z\simeq6$--7 Lyman-break galaxies with $M_{\textrm{UV}}<-20$. These observed galaxies have a wide range of sizes, consistent with the sizes of the general \BlueTides galaxy sample.
The most massive black hole hosts in \BlueTides have small radii of $R_{0.5}=0.41\substack{+0.18 \\ -0.14}$ kpc, as do the quasar hosts, with $R_{0.5}=0.40\substack{+0.11 \\ -0.09}$ kpc.
The total sample of bright black holes ($L_{\textrm{AGN}}>10^{45.1} \textrm{erg s}^{-1}$) has a wider distribution of galaxy sizes with a larger median value, $R_{0.5}=0.51\substack{+0.27 \\ -0.21}$ kpc,
as do hidden quasars, which have $R_{0.5}=0.56\substack{+0.30 \\ -0.23}$ kpc. For comparison, galaxies with similar masses and luminosities ($M_{\textrm{Halo}}>10^{11.3} M_\odot$, $M_\ast>10^{9.5} M_\odot$, $M_{\textrm{UV,Host~(dust)}}<-20.5$) have sizes of $R_{0.5}=0.71\substack{+0.28 \\ -0.25}$ kpc.
A feature of massive black hole and quasar hosts is therefore that they are generally very compact, and are some of the smallest galaxies present at $z=7$.
This is consistent with the lower redshift conclusions of \citet{Bornancini2020}, who observe that $1.4\leq z\leq2.5$ AGN and quasar hosts are more compact than star-forming galaxies.
\citet{Silverman2019} find that $1.2\leq z\leq1.7$ AGN hosts have intermediate sizes, between those of star-forming, disc-dominated galaxies, and more compact quiescent, bulge-dominated galaxies.

Observations of high-redshift quasar host galaxies at sub-mm wavelengths generally measure larger extents of their gas and dust distributions than our predictions for the sizes of their stellar distributions.
For example, studies of [CII] line emission in $z\gtrsim6$ quasars using ALMA find radii of 1.7--3.5 kpc \citep{Wang2013}, 2.1--4.0 kpc \citet{Venemans2015}, a median radius of 2.25 kpc \citep{Willott2017}, and for low-luminosity quasars, radii of 2.6--5.2 kpc \citep{Izumi2018} and 2.1--4.0 kpc \citep{Izumi2019}.
In a larger study, \citet{Decarli2018} measured the [CII] emission of a sample of 27 $z>5.94$ quasars with ALMA, finding radii of 1.2--4.1 kpc. 
These sizes are much larger than our predictions for the stellar distributions of quasar host galaxies, of $R_{0.5}=0.40\substack{+0.11 \\ -0.09}$ kpc, and even the general galaxy distribution of $R_{0.5}=0.71\substack{+0.28 \\ -0.25}$ kpc.
Note, however, that FIR continuum and [CII] line observations generally find that [CII] emission is more extended than the dust continuum emission \citep{Wang2013,Venemans2015,Willott2017,Izumi2018,Izumi2019}, with in some cases the continuum radius smaller than the [CII] radius by even a factor of $\sim3$ \citep{Izumi2019}. As the different observations trace different components, this suggests that the gas and dust, and likewise potentially the stars in galaxies, may follow different distributions \citep[see also][]{Khandai2012,Lupi2019}. 

Using the FIRE simulation and a radiative transfer code to study galaxies in $1<z<5$, \citet{Cochrane2019} find that emission from the stellar component is generally less extended than that of dust continuum emission, cool gas and dust; an example galaxy is quoted as having $R_\ast\simeq0.8$ kpc, $R_{\textrm{cool~gas}}\sim2.7$ kpc, and $R_{\textrm{dust}}\sim2.4$ kpc.
Observations of main-sequence star-forming galaxies at $z\simeq4$--6 with both ALMA and HST find that the [CII] radii exceed the rest-frame UV radii by factors of $\sim2$--3, with median sizes $R_{[CII]}=2.1\pm0.16$ kpc, $R_{\mathrm{F160W}}=0.91\pm0.06$ kpc and $R_{\mathrm{F814W}}=0.66\pm0.04$ kpc, where $R_{\mathrm{F160W}}$ and $R_{\mathrm{F814W}}$ are radii measured from the HST F160W and F814W filters, respectively \citep{Fujimoto2020}.
Thus, it seems reasonable that while ALMA observations find extended emission in quasar hosts from dust and cold gas, our predictions expect the stellar emission to be much more compact. 
We will use \BlueTides to make predictions for the gas and dust properties of quasar hosts and compare these to the stellar properties in future work.

\section{Quasar observations}
\label{sec:Observations}
\begin{figure}
\begin{center}
\includegraphics[scale=0.9]{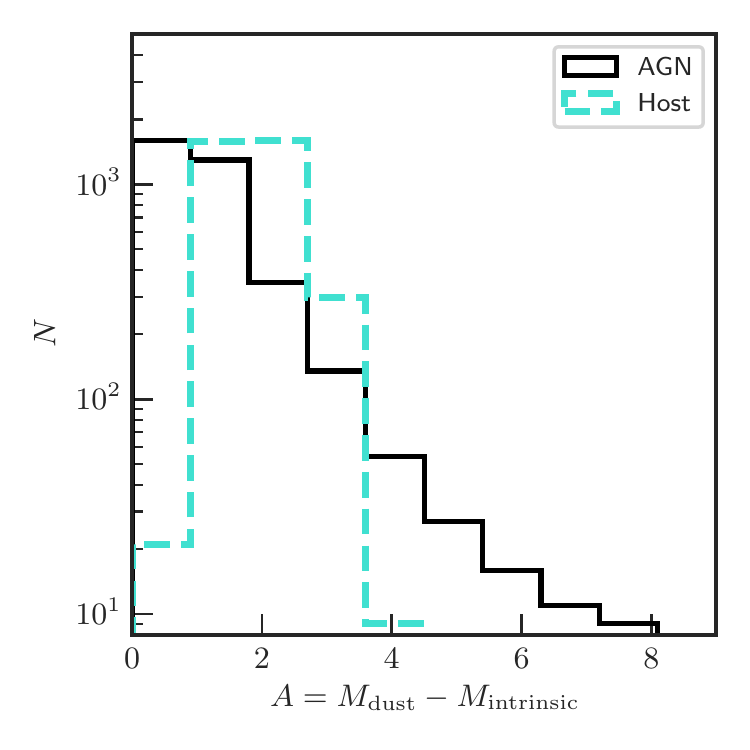}
\caption{The distributions of dust extinction $A=M_{\textrm{dust}}-M_{\textrm{intrinsic}}$ applied to galaxies and AGN. We take the dust extinction for the AGN as that along the line of sight with the least dust-extinction, which is typically the face-on direction, as an optimistic assumption.
}
\label{dust_hists}
\end{center}
\end{figure}
\begin{figure*}
\begin{center}
\includegraphics[scale=0.9]{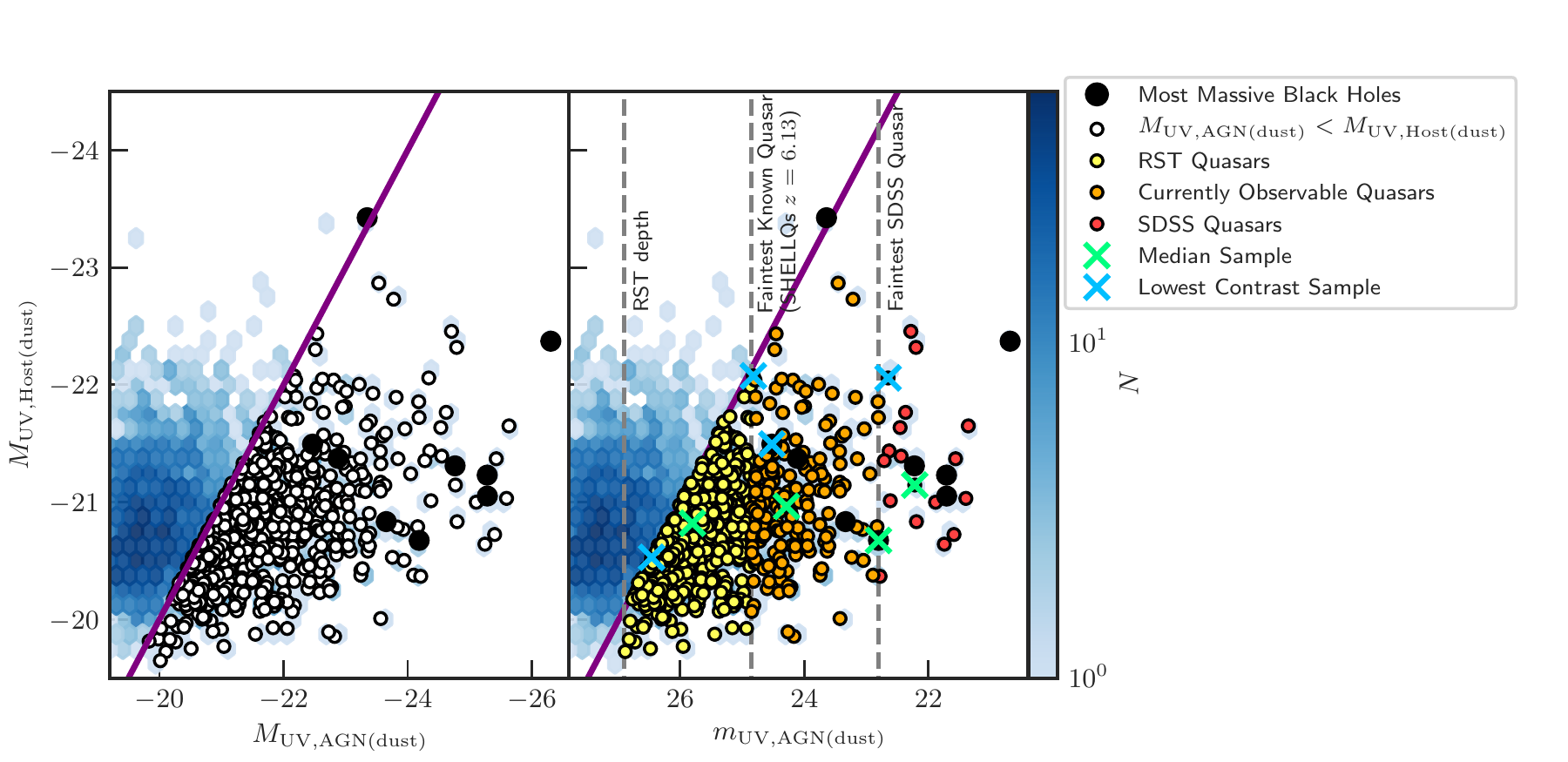}
\caption{The distribution of galaxy and AGN dust-attenuated UV magnitudes for \BlueTides galaxies at $z=7$ (blue density plot). 
The left panel shows the absolute host and AGN magnitudes, and the right shows the apparent AGN magnitude for comparison with observational detection limits.
Using the left panel we classify quasars as those with $M_{\textrm{UV,AGN (dust)}}<M_{\textrm{UV,Host (dust)}}$, since the AGN outshines the host galaxy.
Using the observational detection limits shown in the right panel (grey dashed lines), we further split the quasars into three samples: SDSS quasars, currently observable quasars and RST quasars. We mark with green crosses the median mass black hole in the most massive black hole sample, and the black hole with the median $M_{\textrm{UV,AGN~(dust)}}$ in the SDSS, currently observable, and RST quasar samples (`median sample'). We also mark the black hole in each sample which has the lowest contrast between the quasar and host galaxy brightness (`lowest contrast sample', blue crosses). The most massive black holes are also shown.}
\label{Muv_dust}
\end{center}
\end{figure*}

\begin{figure*}
\begin{center}
\includegraphics[scale=0.9]{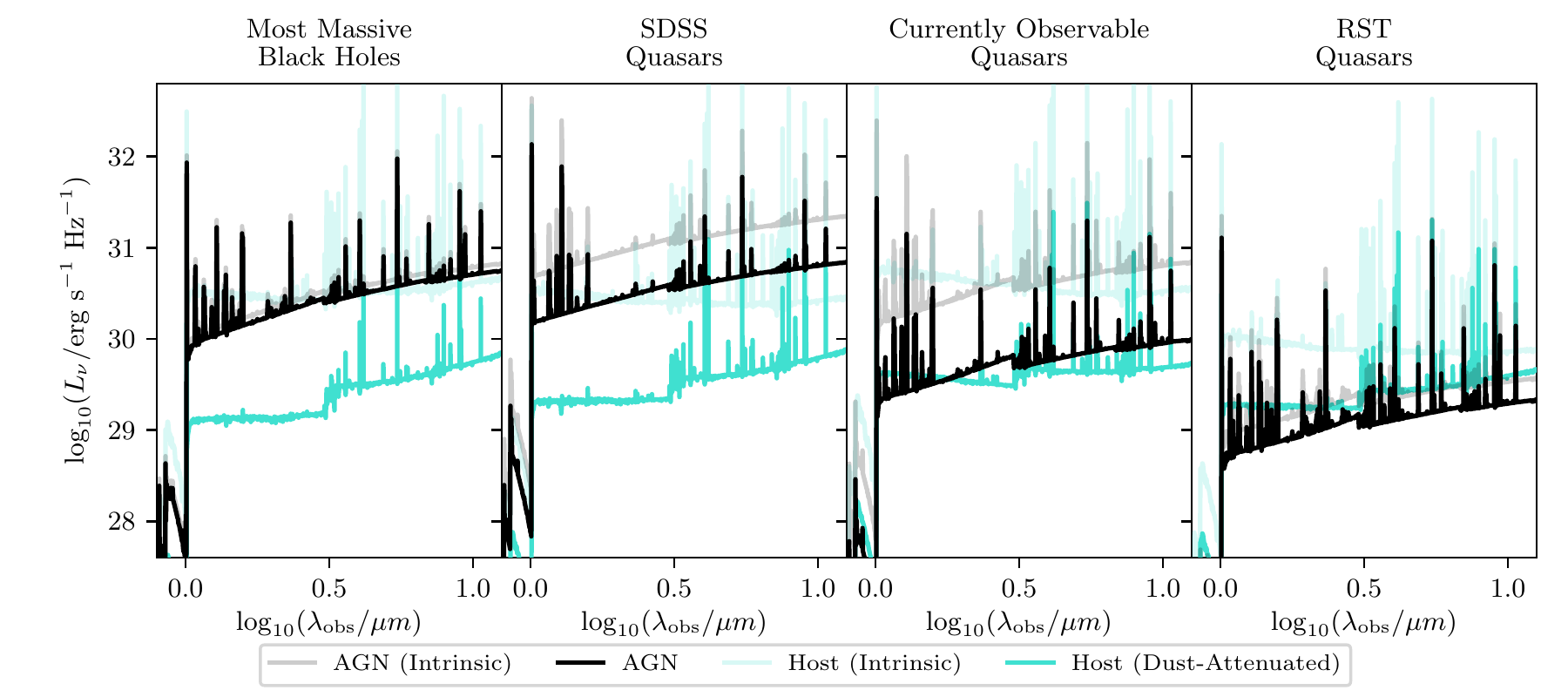}
\caption{The spectra of an AGN and its host galaxy, for the median mass black hole in the most massive black hole sample (`median massive black hole', far left), and the quasar with the median $M_{\textrm{UV,AGN~(dust)}}$ in the SDSS (middle left), currently observable (middle right) and RST (far right) quasar samples (`median quasars').
The upper grey curves show the intrinsic AGN spectra, while the lower black curves shows the AGN spectra with our dust extinction law applied. 
The upper turquoise curves show the intrinsic host galaxy spectra, while the lower turquoise curves show the host galaxy spectra with our dust extinction law applied.
}
\label{SampleSpectra}
\end{center}
\end{figure*}

In Section \ref{sec:intrinsic}, we considered the properties of the host galaxies of the most massive black holes and intrinsically bright quasars in the \BlueTides simulation. We now consider the effects of dust-attenuation and survey magnitude limits to mimic true quasar observations, to make predictions for upcoming observations with JWST.

\subsection{Observable quasar sample selection}
\subsubsection{Magnitude limits of quasar observations}
\label{sec:SelectionLimits}
To select our observable quasar samples, we first consider the magnitude limits of various observational surveys.

The most well-known sample of high-redshift quasars is that of the Sloan Digital Sky Survey \citep[SDSS; e.g][]{Fan2003,Fan2006c,Jiang2016}. The faintest quasar in this sample is SDSS J0129--0035, with $m_{1450}=22.8$, or $M_{1450}=-23.89$ at $z=5.78$ \citep{Wang2013,Banados2016}. This is of similar luminosity to the brightest quasars in the \BlueTides simulation at $z=7$.

The faintest high-redshift quasars observed to date are those discovered in the Subaru High-z Exploration of Low-Luminosity Quasars (SHELLQs) project \citep{Matsuoka2018}, which uses imaging from the Subaru Hyper Suprime-Cam and follow-up spectroscopy using the Gran Telescopio Canarias and the Subaru Telescope. This sample includes $5.7<z<6.8$ quasars down to magnitudes of  $m_{1450}=24.85$, or $M_{1450}=-21.93$ (HSC J1423--0018 at $z=6.13$).

Surveys with upcoming facilities will significantly increase the sample of known high-redshift quasars.
The Euclid spacecraft, expected to launch in the latter half of 2022, will perform a wide survey of 15,000 square degrees to a magnitude of 24.0 in the $Y$, $J$ and $H$ bands, and a deep survey of 40 square degrees to a magnitude of 26.0 \citep{Laureijs2011}.
The Vera C. Rubin Observatory will perform the Legacy Survey of Space and Time (LSST), a survey over 18,000 square degrees reaching depths of magnitude 26.1 in the $z$- band and $24.9$ in the $y$-band, commencing in 2023 \citep{LSST2009}.
These large surveys will discover a large number of high-redshift quasars, complementing the smaller, existing sample of SHELLQs quasars, which found quasars to a similar depth. 
The deep Euclid survey will discover even fainter quasars, although its much smaller area will result in a smaller quasar sample.

At the forefront of upcoming high-redshift quasar discovery surveys is the RST High Latitude Survey, which will cover 2,000 square degrees to a magnitude of 26.9 in the $Y$, $J$ and $H$ bands \citep{Spergel2015}. While RST will not launch until at least 2025, potentially beyond the 5-year mission plan of JWST, the large volume and significant depth of this survey will result in the largest sample of faint quasars in the forseeable future.
We assume that there will be some bluer comparison data of significant depth which can be used to select dropouts, so take $m_{1450}=26.9$ as the faintest $z=7$ AGN luminosity that could be detected by RST. 
%https://smd-prod.s3.amazonaws.com/science-red/s3fs-public/atoms/files/Kruk_APAC__July2018.pdf

\subsubsection{Observable quasars}
In Section \ref{sec:intrinsic} we selected our quasar sample based on the intrinsic UV-band magnitudes of the black holes and host galaxies (Figure \ref{Muv_gal_agn}), with `quasars' defined as black holes which had intrinsic magnitudes brighter than their hosts. In Figure \ref{dust_hists} we show the difference between dust-attenuated and intrinsic magnitudes for both AGN and host galaxies, $A=M_{\textrm{dust}}-M_{\textrm{intrinsic}}$, as calculated following the procedures outlined in Section \ref{sec:dust}. 
Here, and throughout the remainder of this paper, we take the dust extinction for the AGN as that along the line-of-sight with the minimum $\tau_{\rm UV, AGN}$, as an optimistic estimate of the AGN dust extinction. This generally corresponds to the face-on direction \citep[see][]{Ni2019}.

Figure \ref{dust_hists} shows that AGN and host galaxies experience a similar level of dust attenuation in the majority of cases. The AGN population, however, exhibits a small tail in the distribution extending to large $A$. These AGN with extreme dust attenuation are a mixed population of black holes, with a variety of masses and accretion rates. This extinction results in some of the `intrinsic quasars' having dust-attenuated AGN magnitudes that no longer outshine their host galaxy. 
It is therefore important when making mock observational samples to select them based on their dust-attenuated magnitudes.

In Figure \ref{Muv_dust} we show the relation between galaxy and AGN dust-attenuated UV magnitudes for the \BlueTides galaxies.
Since surveys are limited by apparent and not absolute magnitude, we convert our AGN magnitudes using $m-M=\textrm{Distance Modulus} -2.5\log(1+z)=46.99$.
This uses a k-correction of $2.5\log(1+z)$, which accounts for the flux per unit wavelength changing with redshift by a factor of $(1+z)$ \citep{Oke1968,Hogg2002}.
We show these AGN apparent magnitudes in Figure \ref{Muv_dust}, and overplot the observational selection limits described in \ref{sec:SelectionLimits}. 

As in Section \ref{sec:intrinsic}, we make the simple assumption that AGN which outshine their host galaxy in the UV-band are classified as `quasars'. In contrast to Section \ref{sec:intrinsic}, however, we perform this classification using dust-attenuated magnitudes: `quasars' are black holes with $M_{\textrm{UV,AGN (dust)}}<M_{\textrm{UV,Host (dust)}}$.
Using the limiting magnitudes from SDSS, SHELLQs and RST, we define our three observable quasar samples as:
\begin{itemize}
\item SDSS quasars: 
$M_{\textrm{UV,AGN (dust)}}<M_{\textrm{UV,Host (dust)}}$ and $m_{\textrm{UV,AGN (dust)}}<22.8$
\item Currently observable quasars: 
$M_{\textrm{UV,AGN (dust)}}<M_{\textrm{UV,Host (dust)}}$ and $22.8<m_{\textrm{UV,AGN (dust)}}<24.85$
\\
\item RST quasars: $M_{\textrm{UV,AGN (dust)}}<M_{\textrm{UV,Host (dust)}}$ and $24.85<m_{\textrm{UV,AGN (dust)}}<26.9$
\end{itemize}

The `SDSS', `currently observable' and `RST' quasar samples contain 23, 177 and 498 quasars respectively, which is all black holes in the simulation with $m_{\textrm{UV,AGN (dust)}}<22.8$, 99 per cent of black holes with $22.8<m_{\textrm{UV,AGN (dust)}}<24.85$, and 38 per cent of black holes with $24.85<m_{\textrm{UV,AGN (dust)}}<26.9$ (see Figure \ref{Muv_gal_agn}). Defining the observable quasar samples using the dust attenuated magnitudes therefore selects a different, larger sample of black holes.

We define the black hole with the median $M_{\textrm{BH}}$ in the most massive black hole sample as the `median massive black hole', and the quasars with the median $M_{\textrm{UV,AGN~(dust)}}$ from each of these three quasar samples as the `median quasars'. We show the spectra of these objects in Figure \ref{SampleSpectra}, as examples.

\subsection{Images of quasar hosts}
\begin{figure}
\includegraphics[scale=0.7]{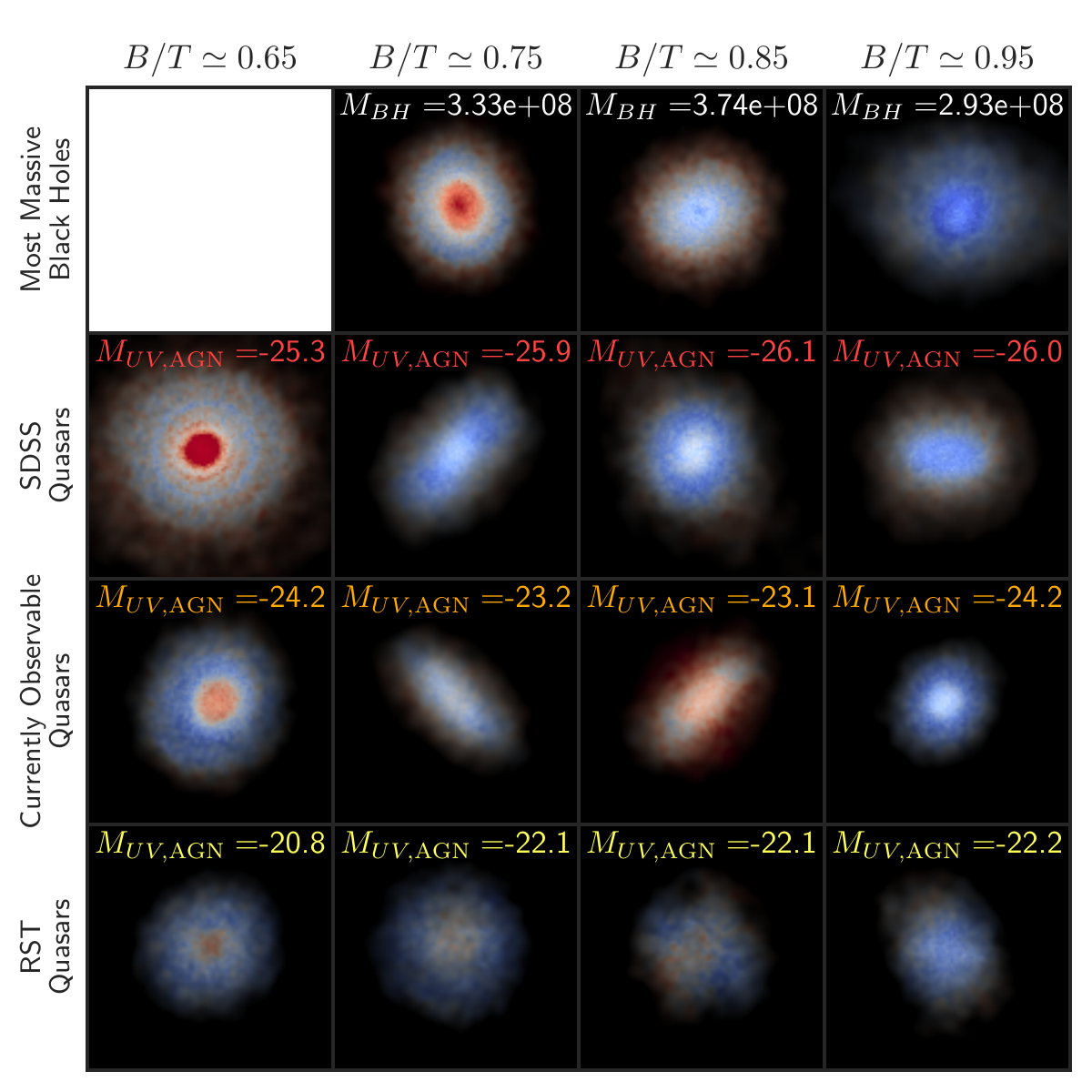}
\caption{The stellar mass distribution of four \BlueTides galaxies from each of the samples: the 10 most massive black holes, SDSS quasars, currently observable quasars, and RST quasars, selected to show a range of morphologies ($B/T\simeq0.65,0.75,0.85$ and 0.95; left to right shows more disc-like to more bulge-dominated galaxies). The minimum $B/T$ for the most massive black hole sample is $\sim0.70$ and so only three galaxies from that sample are shown.
Each galaxy is viewed face-on, with a field-of-view of $3\times3$ kpc. The colour depicts the age of the stellar population, from bluest ($\leq20$ Myr) to reddest ($\geq220$ Myr), with a linear scale.}
\label{BHsamples_faceon}
\end{figure}

\begin{figure}
\includegraphics[scale=0.7]{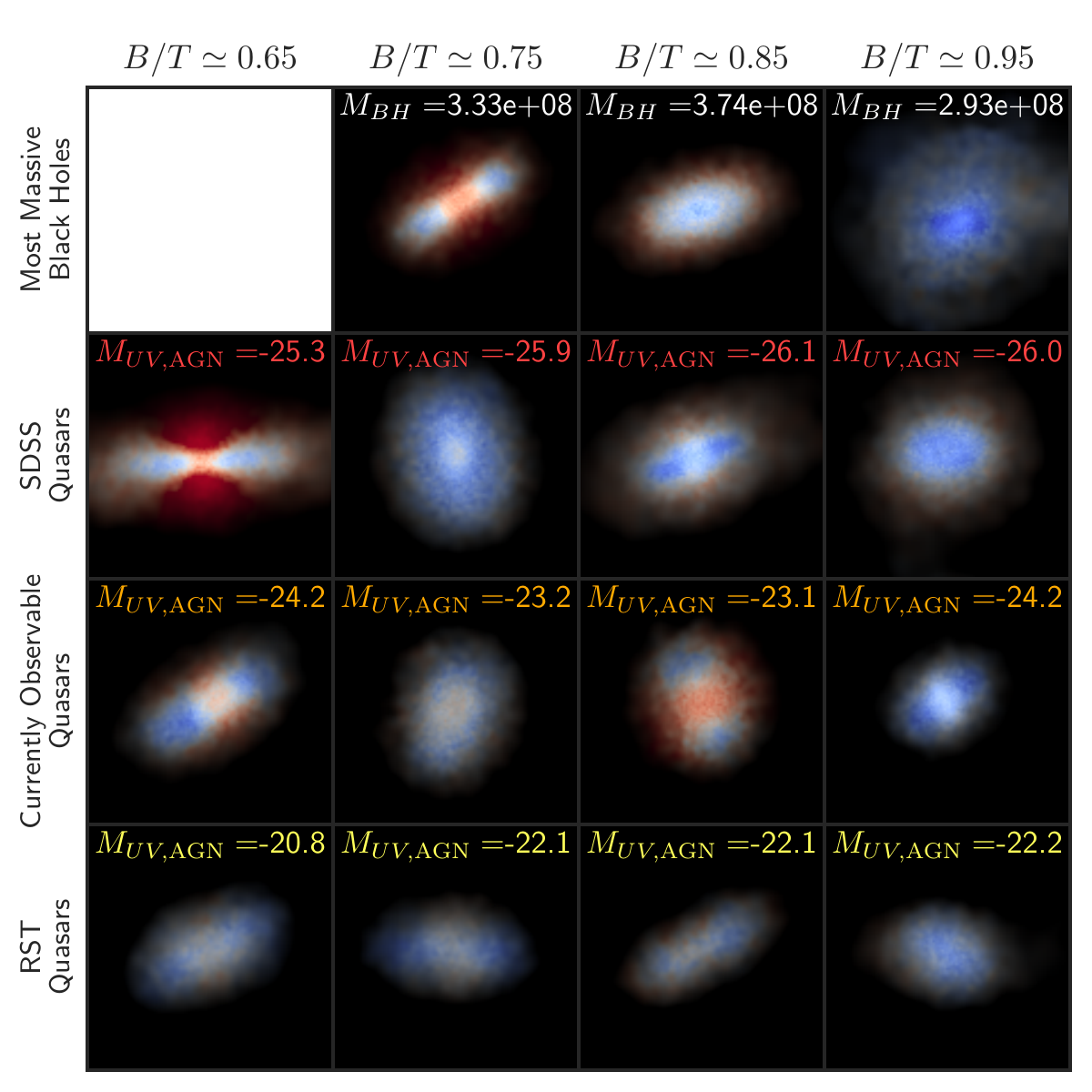}
\caption{The stellar mass distribution of four \BlueTides galaxies from each of the samples: the 10 most massive black holes, SDSS quasars, currently observable quasars, and RST quasars, selected to show a range of morphologies ($B/T\simeq0.65,0.75,0.85$ and 0.95; left to right shows more disc-like to more bulge-dominated galaxies). The minimum $B/T$ for the most massive black hole sample is $\sim0.70$ and so only three galaxies from that sample are shown. Each galaxy is viewed edge-on, with a field-of-view of $3\times3$ kpc.  The colour depicts the age of the stellar population, from bluest ($\leq20$ Myr) to reddest ($\geq220$ Myr), with a linear scale.}
\label{BHsamples_edgeon}
\end{figure}

To visualize the host galaxies of the most massive black holes and `SDSS', `currently observable' and `RST' quasars, we first make images of their mass distributions using \textsc{Gaepsi2},\footnote{\url{https://github.com/rainwoodman/gaepsi2}} a  suite of routines for visualizing SPH simulations. 
We select four galaxies from each sample with a representative range of morphologies ($B/T\simeq0.65,0.75,0.85$ and 0.95) to image. We note that the minimum $B/T$ for the most massive black hole sample is $\simeq0.70$ and so we only select three galaxies from that sample.
The mass distributions of these sample galaxies from a face-on and edge-on perspective are shown in Figures \ref{BHsamples_faceon} and \ref{BHsamples_edgeon} respectively, with colours depicting stellar age. These images show a variety of sizes, shapes and ages of the black hole and quasar host galaxies.

\begin{figure}
\begin{center}
\includegraphics[scale=0.7]{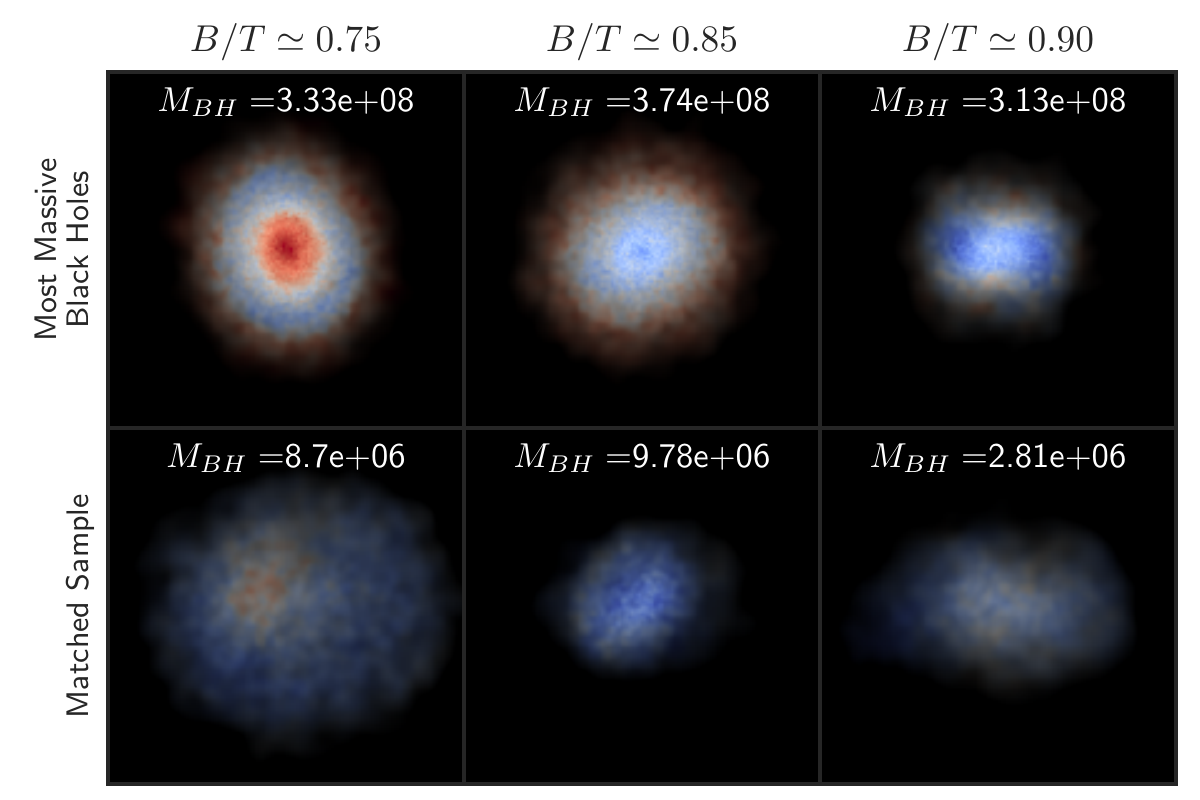}
\caption{The stellar mass distribution of three \BlueTides galaxies from the most massive black hole sample, alongside a matched sample of galaxies with a similar stellar mass and bulge-to-total ratio, but low black hole masses. Each galaxy is viewed face-on, with a field-of-view of $3\times3$ kpc. The colour depicts the age of the stellar population, from bluest ($\leq20$ Myr) to reddest ($\geq220$ Myr), with a linear scale.}
\label{matched_sample}
\end{center}
\end{figure}

We also construct a matched sample for comparison with the three representative most massive black hole hosts. These matched galaxies are chosen to have small black holes, $10^{6.5}<M_{BH}/M_\odot<10^{7}$, but the most similar stellar mass and $B/T$ to those of the three most massive black hole hosts. Images of these representative most massive black holes and the matched galaxy sample can be seen in Figure \ref{matched_sample}. The galaxies in the matched sample are more diffuse and slightly more extended than the hosts of the most massive black holes. This is consistent with our findings from Figure \ref{fig:Size}, which shows that the hosts of massive black holes are more likely to be compact than other galaxies of equivalent stellar mass.

We produce mock JWST images using \textsc{SynthObs},\footnote{\url{https://github.com/stephenmwilkins/SynthObs}} a package for producing synthetic observations from SPH simulations. 
\textsc{SynthObs} takes the flux of each stellar particle, applies the specified photometric filter, and convolves this emission with the corresponding JWST point-spread function (PSF). We assume that the quasar emission comes from a single point, with the quasar thus appearing as a point source convolved with the PSF in the images. By applying the appropriate smoothing, \textsc{SynthObs} produces a mock image with the pixel scale of the instrument. 
We include the effects of noise by adding a random background noise map to the \textsc{SynthObs} images, with noise $\sigma$ from the predicted $10\sigma$ sensitivity of JWST, using a circular photometric aperture 2.5 pixels in radius \citep{NIRCam2017}.
Dust-attenuation is applied following the methodology described in Section \ref{sec:dust}, and we use the minimum AGN dust attenuation of the various sight-lines.

\begin{figure*}
\begin{center}
\includegraphics[scale=0.8]{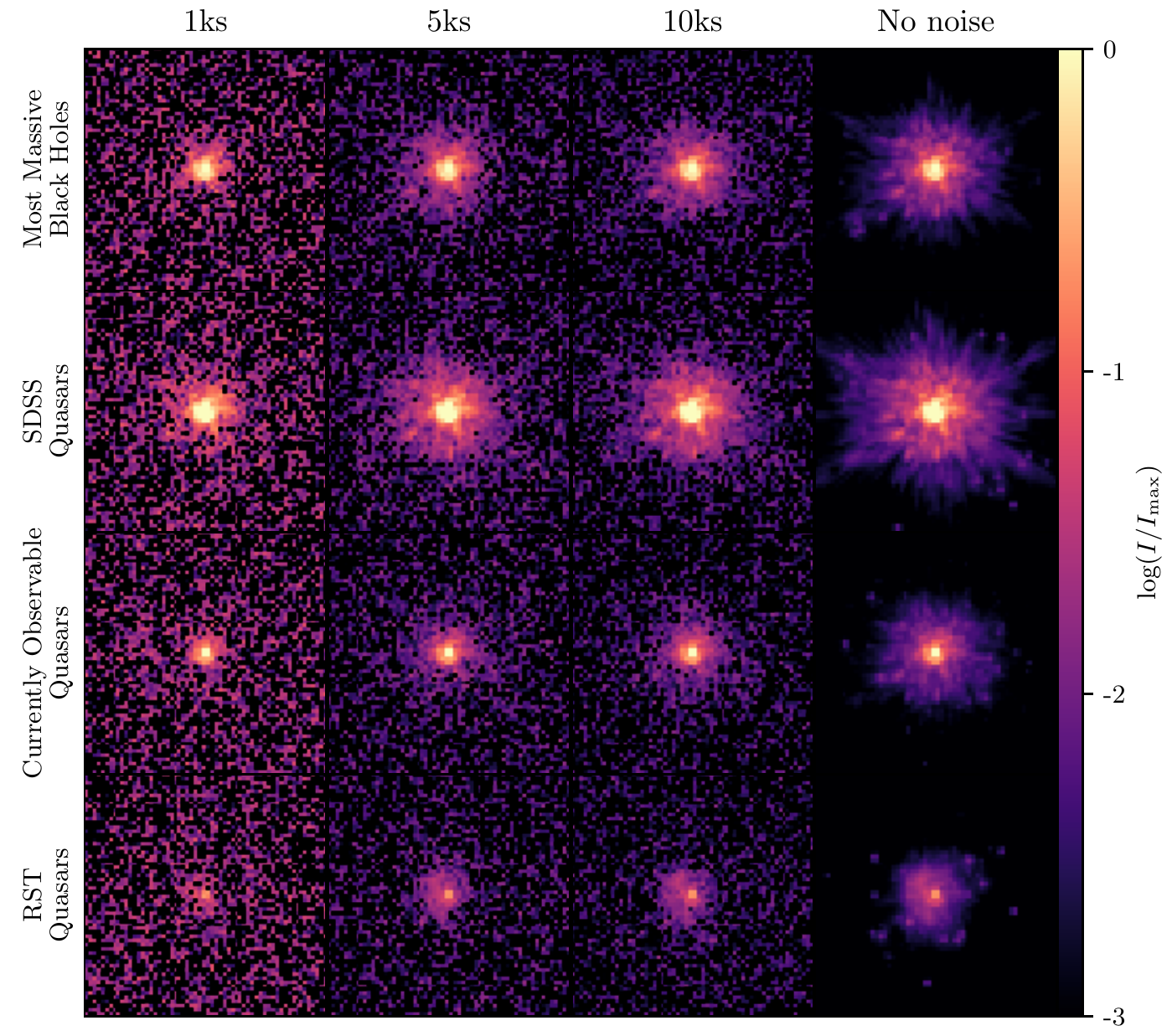}
\caption{Simulated face-on images of \BlueTides galaxies in the JWST NIRCam F200W filter, showing the median massive black hole and the median SDSS, currently observable, and RST quasars (i.e. the galaxies whose spectra is shown in Figure \ref{SampleSpectra}).
The combined quasar and host galaxy emission including dust-attenuation is shown.
In the first three panels from left to right, the images assume an exposure time of 1ks, 5ks, and 10ks, which are predicted to achieve a $10\sigma$ detection of 63.4, 18.2 and 13.2 nJy point sources, using a circular photometric aperture 2.5 pixels in radius \citep{NIRCam2017}.
The right-most panel shows the images with no noise background.
The field-of-view is $10\times10$ kpc, or $1\farcs86 \times 1\farcs86$. 
Note that all panels are shown with the same intensity scale.}
\label{NoiseComparison}
\end{center}
\end{figure*}

In Figure \ref{NoiseComparison} we show mock JWST imaging in the NIRCam F200W filter of the galaxies hosting the median massive black hole and the median quasars (i.e. those whose spectra are shown in Figure \ref{SampleSpectra}).
These images show the combined (dust-attenuated) quasar and host emission with varying exposure times: 1ks, 5ks, 10ks, and an image with no noise background for comparison.
This shows that deep exposure times of $\gtrsim$ 5ks are required to observe the detailed structure present in the noise-less images, which is likely to be necessary for detecting the underlying host galaxy emission, which generally has low surface brightness and is hidden by the bright quasar emission (see discussion below). We therefore choose to adopt an exposure time of 10ks for our mock images herein. We choose to show the F200W filter as an example, as it has the highest sensitivity of the NIRCam wide-band filters, resulting in the least background noise for a given exposure time.

\begin{figure*}
\begin{center}
\includegraphics[scale=0.8]{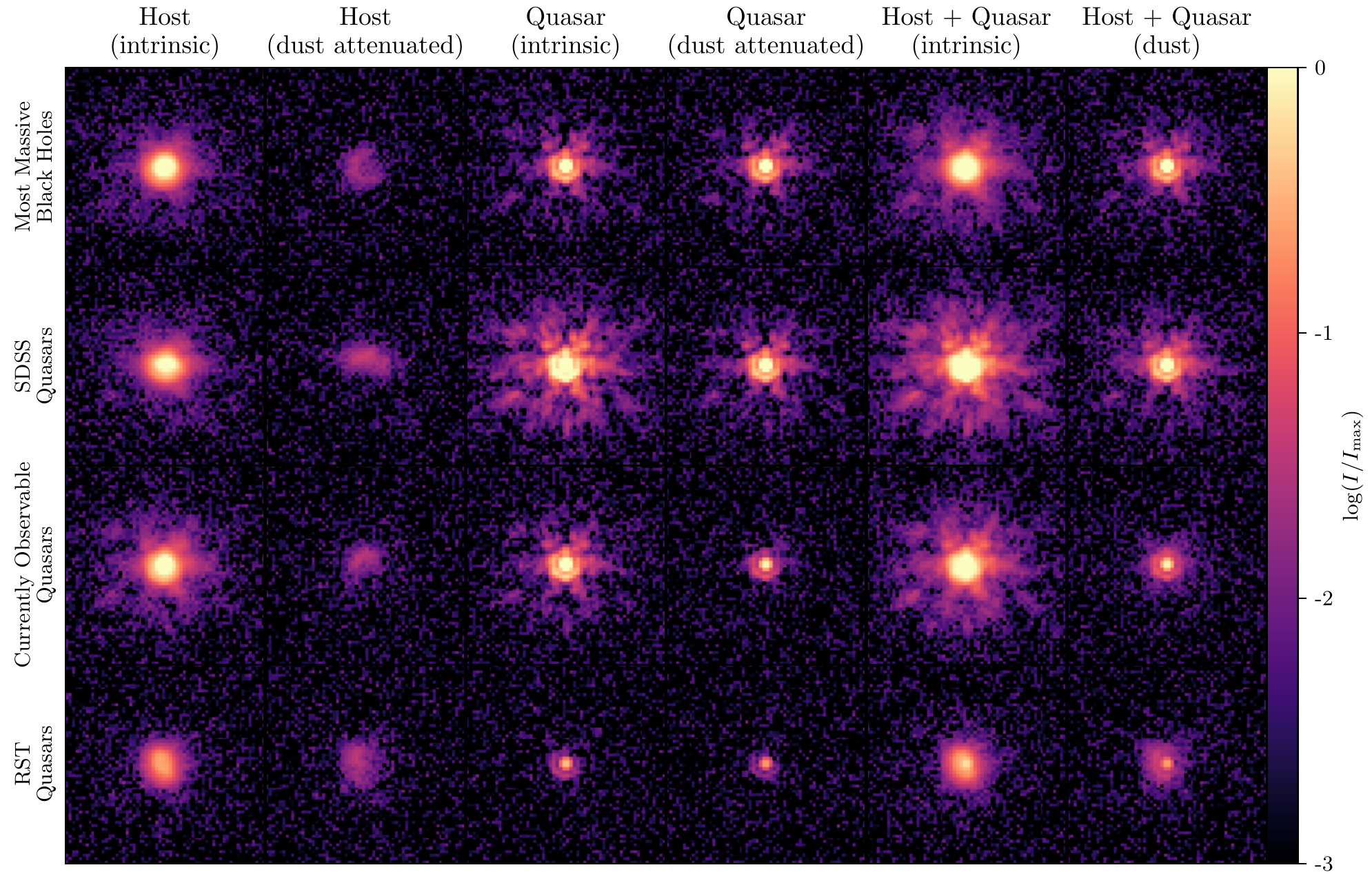}
\caption{Simulated face-on images of \BlueTides galaxies in the JWST NIRCam F200W filter, showing the median massive black hole and the median SDSS, currently observable, and RST quasars (i.e. the galaxies whose spectra is shown in Figure \ref{SampleSpectra}).
The host galaxy emission is shown with and without dust-attenuation in the two left-most panels. The emission from the quasar is shown with and without dust-attenuation in the middle panels, with the combined quasar and host galaxy image shown with and without dust attenuation (applied to both the host and quasar) in the right-most panels. 
These images assume an exposure time of 10ks, which is predicted to achieve a $10\sigma$ detection of 13.2 nJy (AB mag $\sim28.8$) point sources, using a circular photometric aperture 2.5 pixels in radius \citep{NIRCam2017}.
The field-of-view is $10\times10$ kpc, or $1\farcs86 \times 1\farcs86$. 
Note that all panels are shown with the same intensity scale.}
\label{BHsamples_JWST}
\end{center}
\end{figure*}

In Figure \ref{BHsamples_JWST} we show mock JWST imaging in the NIRCam F200W filter of the hosts of the median massive black hole and the median quasars, with and without dust attenuation, and with and without the quasar emission. With a resolution of 0.031 arcseconds, JWST only partially resolves the host galaxies, with diameters of $\sim 0.8$ kpc or $\sim 0.15$ arcseconds at $z=7$. Their emission is centrally concentrated, and so the hosts appear as a smeared PSF at this resolution. However, as the density of dust is highest in the central regions, the dust attenuated images show more interesting, asymmetrical features.

The limited resolution of these small galaxies makes it difficult to distingush the host galaxy once the point-source quasar emission is included in the images.
For the intrinsic images, the image is broader than the quasar image (i.e. the PSF of the telescope), suggesting that an accurate modelling technique should be able to detect the host emission despite the presence of the quasar. However, including the effect of dust-attenuation makes the host more difficult to distinguish from the quasar, as its emission becomes fainter and less extended, particularly for the median massive black hole, SDSS and currently observable quasars.
For these three systems, the brightness contrast between the host and quasar is $\sim1.5$ orders of magnitude at the centre, decreasing with distance from the quasar, with the two having similar brightnesses towards the edge of the host galaxy at $\sim 0.5$''. 
The median RST quasar has a lower contrast between the quasar and host, resulting in the host being more easily visible around the bright, central emission from the quasar. 
Distinguishing the host galaxy from the quasar emission will therefore still be challenging with JWST, even with its improved resolution over HST. 

While it appears that the host galaxies are more easily detected in the fainter quasar samples, this is an effect of the contrast ratio between the host and quasar, and not the quasar's total brightness. For the median massive black hole and the median SDSS, currently observable, and RST quasars, the difference in AGN and host magnitude is 3.51, 3.62, 1.74, and 0.37, respectively (see Figure \ref{Muv_dust}). Thus, for the median massive black hole and SDSS quasar, as the AGN magnitude is much brighter than the host, it completely obscures any host emission. For the median currently observable quasar, the bright point source still dominates, however the total emission is somewhat broader than the quasar PSF, which may allow for a host detection with an accurate modelling technique. For the median RST quasar, the lower contrast ratio results in an easily distinguishable host galaxy.

\begin{figure}
\begin{center}
\includegraphics[scale=0.75]{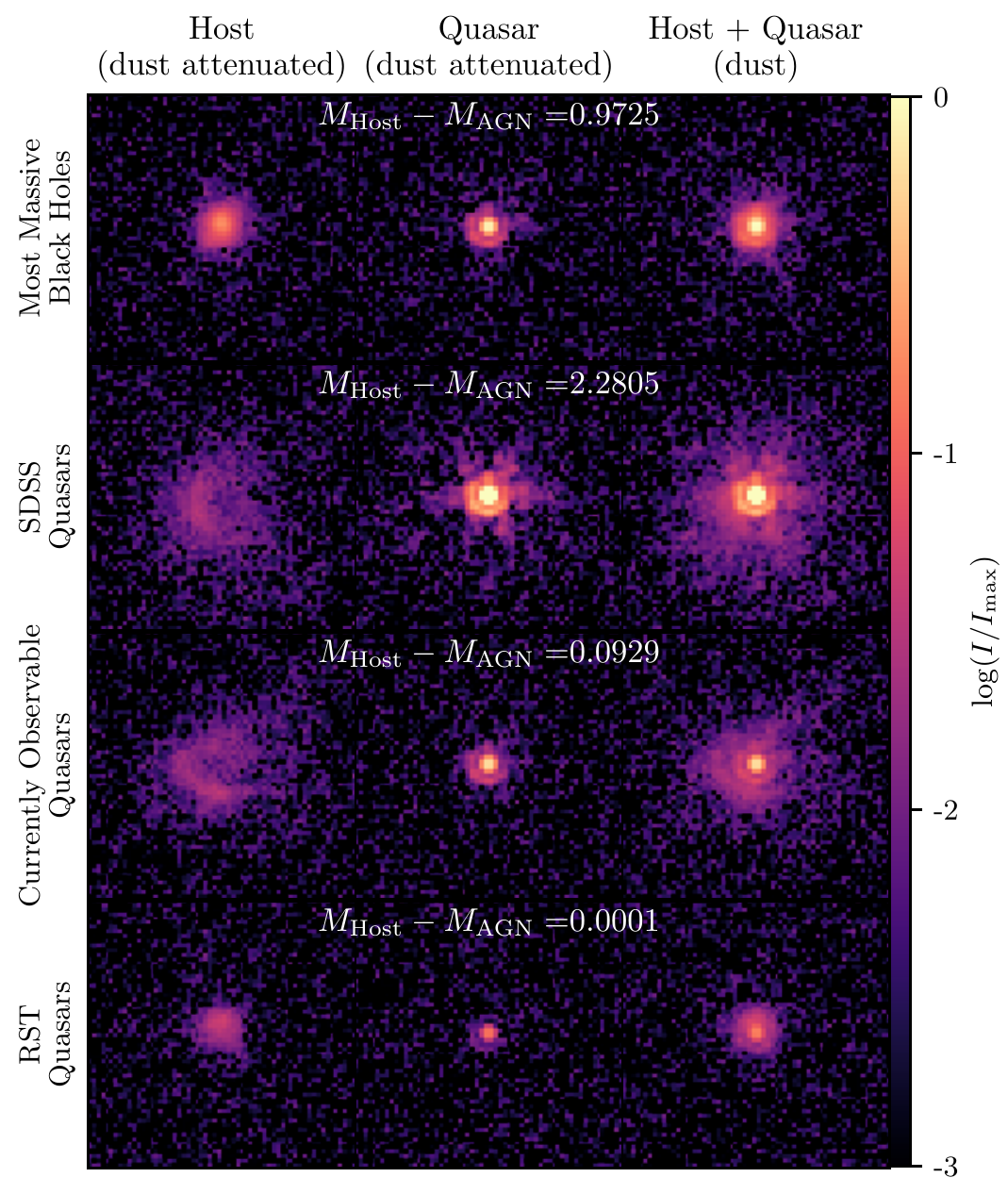}
\caption{Simulated face-on images of \BlueTides galaxies in the JWST NIRCam F200W filter, showing the galaxy from each black hole sample that has the lowest contrast ratio between the quasar and the host $M_{\rm{UV, Host (dust)}}-M_{\rm{UV,AGN (dust)}}$ (with $M_{\textrm{UV,AGN (dust)}}<M_{\textrm{UV,Host (dust)}}$).
The host galaxy emission is shown in the left-most panels, with the emission from the quasar shown in the middle panels. The combined quasar and host galaxy image is shown the right-most panels. All images include the effect of dust-attenuation.
These images assume an exposure time of 10ks, which is predicted to achieve a $10\sigma$ detection of 13.2 nJy (AB mag $\sim28.8$) point sources, using a circular photometric aperture 2.5 pixels in radius \citep{NIRCam2017}.
The field-of-view is $10\times10$ kpc, or $1\farcs86 \times 1\farcs86$. Note that all panels are shown with the same intensity scale.}
\label{MostObservable_JWST}
\end{center}
\end{figure}

To investigate this effect further, we consider the black hole in each sample which has the lowest contrast ratio between the AGN and host luminosity, $M_{\rm{UV, Host (dust)}}-M_{\rm{UV,AGN (dust)}}$ (with $M_{\rm{UV, AGN (dust)}}<M_{\rm{UV,Host (dust)}}$). Mock JWST images of these systems are shown in Figure \ref{MostObservable_JWST}. The host galaxies of these black holes are much easier to distinguish from the quasar point source emission, particularly for those that have the lowest contrast ratios, and for hosts that have more extended emission. Thus, while the host galaxies of quasars will still be difficult to detect with JWST in general, it should be possible to detect the hosts of quasars with low contrast ratios.

\begin{figure*}
\begin{center}
\includegraphics[scale=0.75]{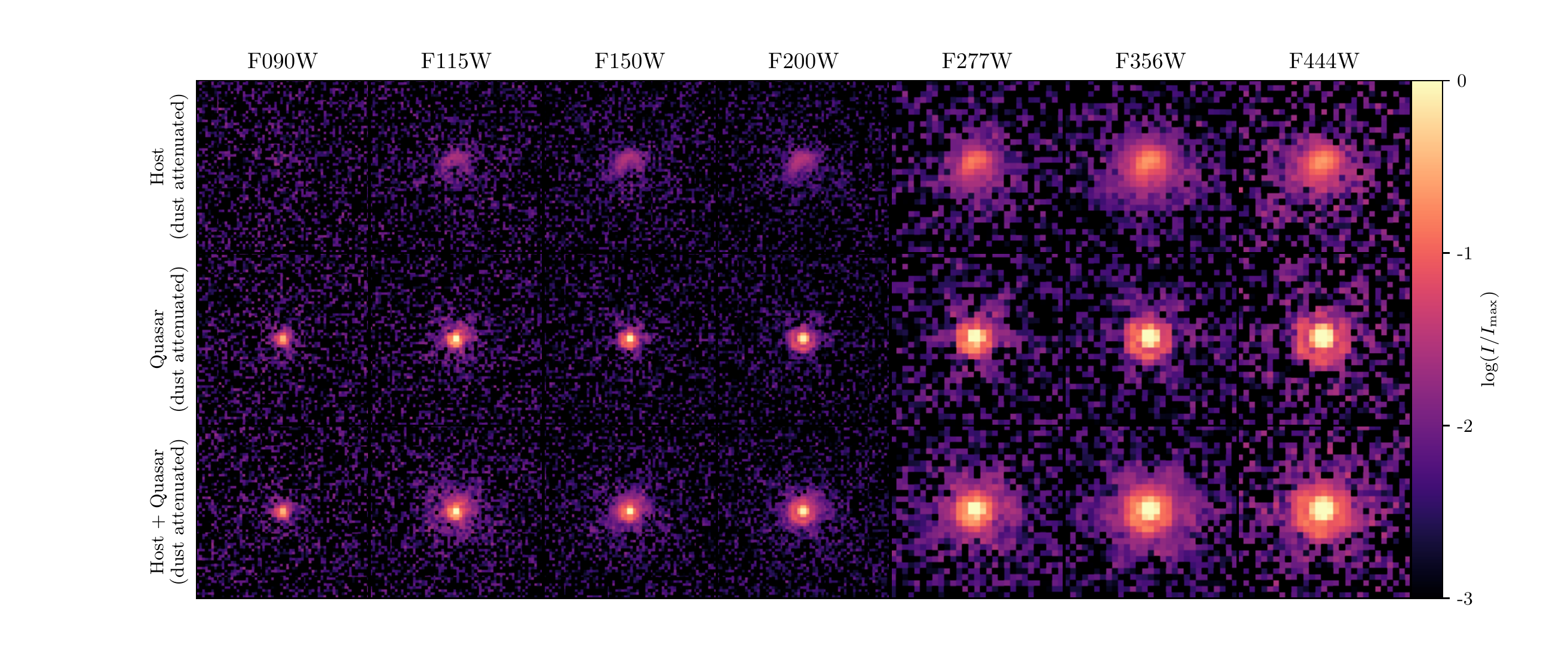}
\caption{Simulated face-on images of the median currently observable quasar in the JWST NIRCam wide-band filters red-ward of the $z=7$ Lyman-break.
The host galaxy emission is shown in the top panels, the emission from the quasar in the middle panels, with the combined quasar and host galaxy image shown in the bottom panels. All images include dust extinction of both the quasar and the host galaxy.
These images assume an exposure time of 10ks, with $10\sigma$ detection sensitivities as predicted by \citet{NIRCam2017}.
The field-of-view is $12\times12$ kpc, or $2\farcs23 \times 2\farcs23$. Note that all panels are shown with the same intensity scale. }
\label{JWST_filters}
\end{center}
\end{figure*}

We show mock JWST images of the median currently observable quasar in all of the NIRCam wide-band filters red-ward of the Lyman-break in Figure \ref{JWST_filters}. 
Most filters show a combined image that is slightly broader than the quasar PSF, however no filter makes the host clearly more detectable.
As the wavelength increases, the resolution of the telescope decreases. Thus, while the contrast ratio of the quasar and its host should be lower at larger wavelengths due to the spectral shapes of quasars and host galaxies (see Figure \ref{SampleSpectra}), redder NIRCam filters do not particularly make the host more easily distinguishable.
The instrument sensitivity increases from the F090W to F200W filters, and then decreases for the higher wavelength filters \citep{NIRCam2017}. The highest sensitivity F200W filter results in the clearest image of the quasar system for a given exposure time (here 10 ks), and thus may offer the best results for detecting quasar host galaxies with JWST.

These conclusions are based only on examining the resulting images by eye. In future work, we will make more detailed and robust predictions for the detectability of quasar hosts with JWST by running an observational technique used to detect quasar host galaxies on these simulated images.

\section{Biases in the observed scaling relations}
\label{sec:Relation}
\begin{figure*}
\begin{center}
\includegraphics[scale=0.9]{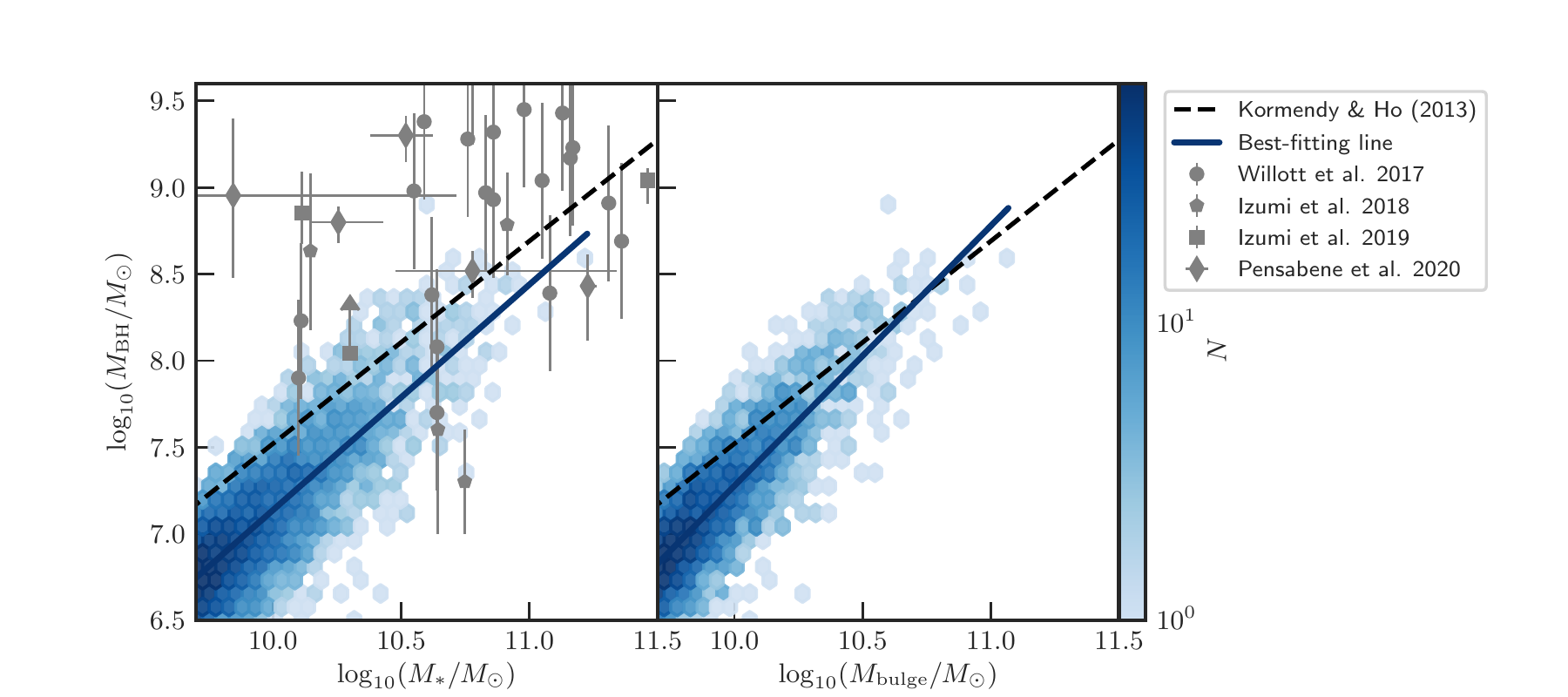}
\caption{The relation between black hole mass and stellar mass (\textit{left}) and black hole mass and bulge mass (\textit{right}) for \BlueTides galaxies at $z=7$, and their best-fitting relations as given in Equations \ref{MBH_M_Relation} and \ref{MBH_Mbulge_Relation}.
We plot a range of observations of $5\lesssim z \lesssim7$ quasars from the literature \citep{Willott2017,Izumi2018,Izumi2019,Pensabene2020}, assuming their stellar mass is equal to their measured dynamical mass.
We also plot the observed black hole--bulge mass relation at $z=0$ \citep{Kormendy2013}. This relation is also shown in the left (stellar mass) panel for comparison, assuming that the hosts are pure elliptical galaxies with $M_\ast=M_{\textrm{bulge}}$.}
\label{BHStellarBulgeMass}
\end{center}
\end{figure*}

We now consider the $z=7$ black hole--stellar mass and black hole--bulge mass relations predicted by \BlueTidesns, shown in Figure
\ref{BHStellarBulgeMass}. 
The best-fitting relations for black holes with $M_{\textrm{BH}}>10^{6.5}M_\odot$ and galaxies with $M_\ast>10^{9.7}M_\odot$ are
\begin{equation}
\label{MBH_M_Relation}
\log(M_{\textrm{BH}}/M_\odot)=(1.30\pm 0.02) \log(M_\ast/M_\odot)-(5.9\pm 0.2),
\end{equation}
and
\begin{equation}
\label{MBH_Mbulge_Relation}
\log(M_{\textrm{BH}}/M_\odot)=(1.50 \pm 0.02 )\log(M_{\textrm{Bulge}}/M_\odot)-(7.7 \pm 0.2),
\end{equation}
with errors calculated from 10,000 bootstrap realisations.
The standard deviation of the residuals, or scatter, is 0.2 dex for both relations, so the simulation shows no preference for a tighter correlation of black hole mass with either total or bulge stellar mass. We note that these relations are unlikely to be sensitive to the black hole seeding prescription, as we consider only black holes that have grown significantly above the seed mass of $10^{5.8}M_\odot$.

The \BlueTides black hole--bulge mass relation at $z=7$ is steeper than the observed local relation \citep{Kormendy2013}, which is also shown in Figure \ref{BHStellarBulgeMass}. However, the simulations and local observations are reasonably consistent, particularly at the highest masses where the observed relation is best measured.
Figure \ref{BHStellarBulgeMass} also shows a range of observations of $5\lesssim z \lesssim7$ quasars, assuming their stellar mass is equal to their measured dynamical mass \citep{Willott2017,Izumi2018,Izumi2019,Pensabene2020}. Quasars observed with $M_{\textrm{BH}}>10^{8.5}M_\odot$ show a wide range of dynamical masses, and generally lie above the local relation. These observed black holes are larger than those present in the \BlueTides simulation at $z=7$, so a comparison cannot be made. Observations of lower-luminosity quasars with $M_{\textrm{BH}}<10^{8.5}M_\odot$ are consistent with the \BlueTides relation.

\begin{figure}
\begin{center}
\includegraphics[scale=0.9]{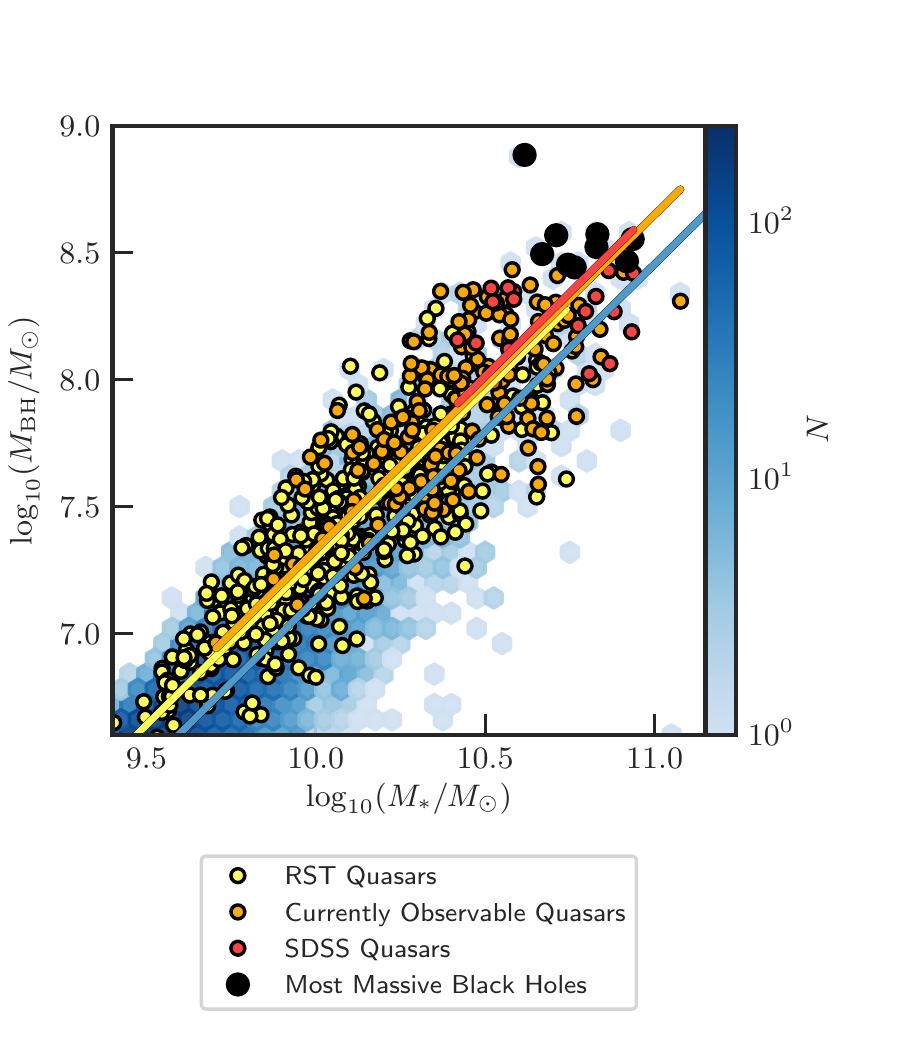}
\caption{The relation between black hole mass and stellar mass. The blue density plot shows the distribution for all \BlueTides galaxies, while the circles show the most massive black holes, and SDSS, currently observable, and RST quasar samples (see legend).
The solid lines are fits to the total (blue), SDSS (red), currently observable (orange), and RST quasar (yellow) samples, constraining the slope to be the same as that for the total sample (Equation \ref{eq:BHStellarMass}). 
}
\label{BHStellarMass}
\end{center}
\end{figure}

In Figure \ref{BHStellarMass} we plot the black hole--stellar mass relation for the most massive black holes, SDSS quasars, currently observable quasars, and RST quasars. This shows that the hosts of the most massive black holes and quasars have large black hole masses for their stellar mass. 
To investigate the effect of this bias on the observed black hole--stellar mass relation we make fits to the SDSS, currently observable, and RST quasar samples, constraining the slope to be equal to that of the total sample:
\begin{equation}
\log(M_{\textrm{BH}}/M_\odot)=1.32 \log(M_\ast/M_\odot)+b.
\label{eq:BHStellarMass}
\end{equation}
where for the full sample $b=-6.06$ (Equation \ref{MBH_M_Relation}). 
The SDSS, currently observable, and RST quasar samples have a normalization of $b=-5.85$, $-5.87$ and $-5.90$, respectively, $\sim$ 0.2 dex higher than that of the full galaxy sample. Quasar samples are therefore biased samples of the intrinsic black hole--stellar mass relations, consistent with expectations from observations \citep[e.g.][]{Lauer2007,Salviander2007,Schulze2014,Willott2017}.
Our simulation provides a calibration of this systematic effect.
A similar bias to larger black hole masses is theoretically expected to occur when observing the black hole--velocity dispersion relation \citep[see e.g.][]{Volonteri2011}.

Interestingly, we find that the fainter quasar samples are as biased as the bright quasar samples in measuring the intrinsic black hole--stellar mass relation.
To investigate this further, we remove the `quasar' constraint that $M_{\textrm{UV,AGN (dust)}}<M_{\textrm{UV,Host (dust)}}$, and instead consider all AGN brighter than each survey limit. We make fits to these SDSS, currently observable, and RST AGN samples, again constraining the slope to be equal to that of the total sample (Equation \ref{eq:BHStellarMass}). The SDSS, currently observable, and RST AGN samples have a normalization of $b=-5.85$, $-5.87$ and $-5.97$, respectively, relative to the total sample with $b=-6.06$. Here, the bias decreases with survey depth. This is qualitatively consistent with the expectations of \citet{Lauer2007}, which found that for a survey limited only by $L_{\textrm{AGN}}$ and not the host properties, the bias will decrease, but is not removed entirely, as the survey goes to fainter luminosity limits.

The constraint that $M_{\textrm{UV,AGN (dust)}}<M_{\textrm{UV,Host (dust)}}$ for quasars therefore results in the RST quasar sample measuring a larger bias in the black hole--host mass relation than for all RST AGN. We consider
black hole samples with $M_{\textrm{UV,AGN (dust)}}-M_{\textrm{UV,Host (dust)}}<-2$, 0 and $2$, with no magnitude limit, and find the best fits to Equation \ref{eq:BHStellarMass}. These fits have normalizations of $b=-5.79$, $-5.89$ and $-5.99$ respectively; the brighter the AGN relative to the host, the larger the black hole mass at fixed stellar mass. Thus, to reduce the bias in the measured black hole--stellar mass relation, observations should target AGN that are faint relative to their host galaxies. There are more of these objects in fainter AGN samples, hence the sample of all RST AGN has a reduced bias compared with the brighter SDSS and currently observable AGN samples.

\section{The environments of high-redshift quasars}
\label{sec:Mergers}
We now study the environments of high-redshift quasars in the \BlueTides simulation, by investigating neighbouring galaxies that host a black hole.
The requirement for a companion to host a black hole is due to no halo sub-finding algorithm being implemented on the simulation; it cannot identify multiple galaxies within an individual dark matter halo, which are precisely the systems we are interested in. As an approximation, we identify neighbouring galaxies via their black holes, and assign all particles within $R_{0.5}$ of the black hole to that galaxy.

\subsection{The number of nearby galaxies}
\label{sec:NumNeighbours}

\begin{figure*}
\includegraphics[scale=1]{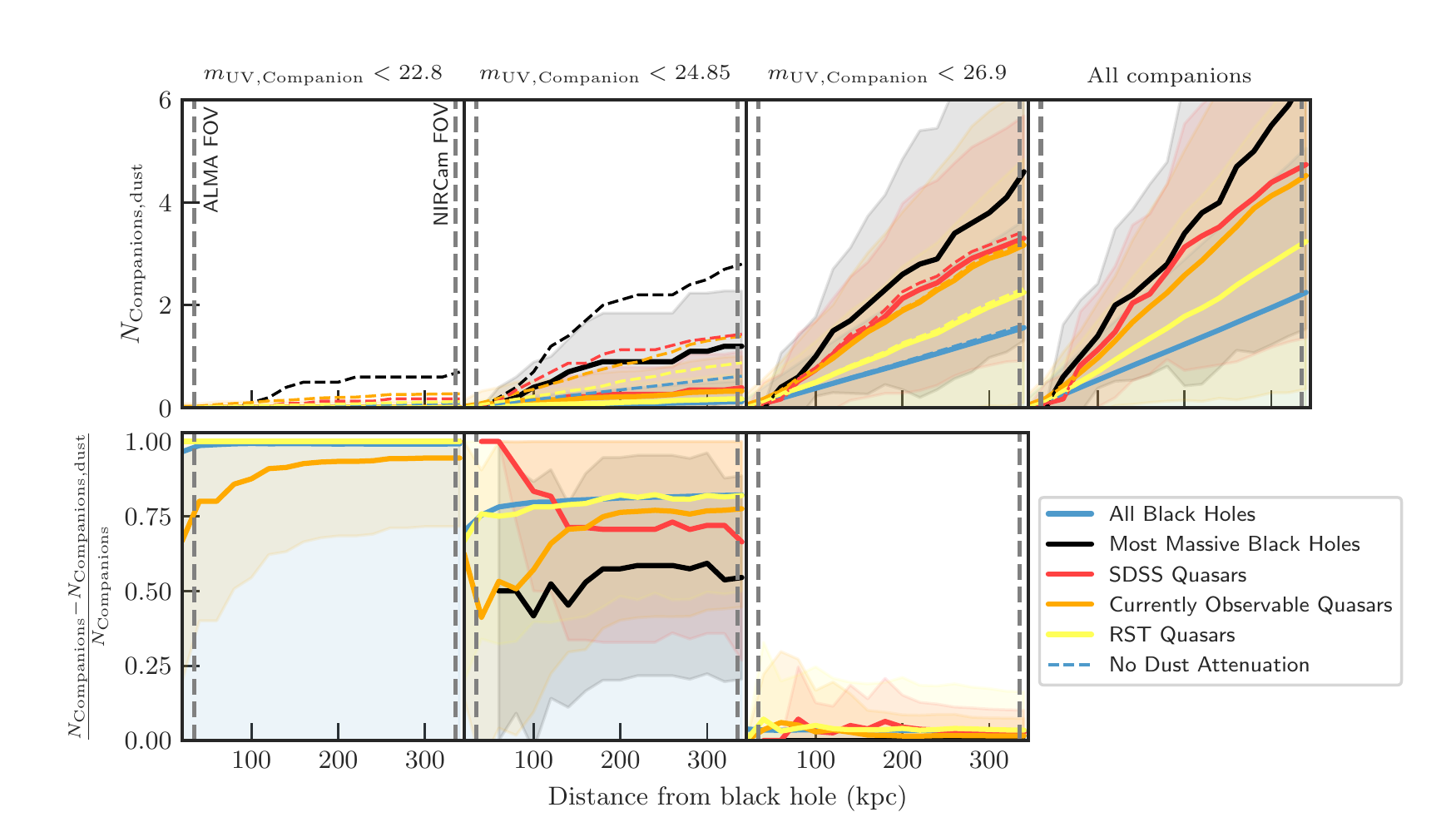}
\begin{center}
\hspace{-1cm}
\caption{ \textit{Top row:} The average number of companion galaxies around each black hole in the sample, as a function of distance from the black hole. 
Each panel shows companions brighter than a given magnitude limit. From left to right: companions with $m_{\textrm{UV}}<22.8$, i.e. the magnitude of the faintest SDSS quasar; companions with $m_{\textrm{UV}}<24.85$, i.e. the magnitude of the faintest currently known quasar; companions observable in RST, with $m_{\textrm{UV}}<26.9$; and all companion galaxies. The solid coloured lines show the average number of companions at each magnitude limit, including the effect of dust attenuation, and shaded regions show the $\pm 1 \sigma$ range.
The dashed coloured lines show the average number of companions that are intrinsically brighter than the magnitude limit (i.e. without the effect of dust attenuation). 
Vertical grey dashed lines show the ALMA \citep{Trakhtenbrot2017} and JWST NIRCam fields of view.\newline
\textit{Bottom row:} The fraction of companions around a black hole that are missed due to dust attenuation, at each magnitude limit. The solid lines show the average fraction, and the shaded regions show the $\pm 1 \sigma$ range. The `all black hole' sample refers to black holes with $M_{\textrm{BH}}>10^{6.5}M_\odot$.}
\label{NumNeighbours}
\end{center}
\end{figure*}

We first consider the number of nearby galaxies around each black hole ($M_{\textrm{BH}}>10^{5.8}M_\odot$). We compare the most massive black hole and quasar samples to the overall sample of black holes with $M_{\textrm{BH}}>10^{6.5}M_\odot$.

In Figure \ref{NumNeighbours} we show the average number of neighbours within a given distance that would be observed at various magnitude limits, for black holes in each sample. 
To a magnitude limit of the faintest SDSS quasar, $m_{\textrm{UV}}=22.8$, no companions are detected within 340 kpc, on average, around any black hole.
At a deeper magnitude limit of $m_{\textrm{UV}}=24.85$, the magnitude of the faintest known high-redshift quasar, no black hole samples are predicted to have nearby companions within $\sim 80$ kpc, on average. The average number of companions increases slightly at larger distances for the massive black hole and quasar samples. At distances $\gtrsim150$ kpc, the most massive black holes have an average of $\sim1$ companion with  $m_{\textrm{UV}}<24.85$, more than expected for the overall sample. However, this enhancement is not significant given the uncertainties.

Many more companions would be observable with RST, with a survey depth of $m_{\textrm{UV}}=26.9$. The average number of $m_{\textrm{UV}}<26.9$ companions for each sample increases with distance from the black hole, with the most massive black holes having an average of $\sim2$ companions within $\sim100$ kpc, and $\sim4$ companions within $\sim300$ kpc. This sample shows the largest number of companions, with the SDSS quasars, currently observable quasars, RST quasars and all black holes having progressively less companion galaxies, with all black holes having an average of $\sim0.5$ companions within $\sim100$ kpc, and $\sim1.5$ companions within $\sim300$ kpc.
A similar enhancement is seen when considering all companion galaxies, with no magnitude limit. On average, the most massive black holes have the most companion galaxies within 50--340 kpc, followed by the quasar samples from brightest to faintest, with the enhancement above the overall black hole sample largest at larger distances ($>150$ kpc). 
As we predict more neighbouring galaxies at larger separations ($>50$ kpc), the majority of companion galaxies are too distant to be detected in the small field of view of ALMA.

The most massive black holes are more likely to be found in denser environments than the typical $M_{\textrm{BH}}>10^{6.5}M_\odot$ black hole, with quasars showing a weaker enhancement. 
However, the increased average number of companions found around the most massive black holes and quasars relative to the general sample is statistically insignificant, with a large variation seen in the number of galaxies around each black hole.
Our conclusions are consistent with the more comprehensive analysis of \citet{Habouzit2019}, who investigated black hole environments in the Horizon-AGN simulation at $z\simeq4$--6. \citet{Habouzit2019} found that, on average, massive black holes live in regions with more nearby galaxies, with an excess of up to 10 galaxies within 1 cMpc at $z\simeq4$--5. The enhancement is larger for more massive black holes. \citet{Habouzit2019} found a diversity in number counts, with some massive black holes having similar numbers of nearby neighbours to the average number counts, consistent with our expectations.

Companion galaxies have been observed near high-redshift quasars, particularly in sub-mm observations \citep[e.g.][]{Wagg2012,McGreer2014,Decarli2017,Willott2017,Neeleman2019}.
In the rest-frame UV/optical, \citet{McGreer2014} detected companion galaxies around two quasars, although found that bright companion galaxies within 20 kpc are uncommon, with an incidence of $\lesssim2/29$ for $\gtrsim 5L^\ast$ galaxies and $\lesssim1/6$ for $2\lesssim L \lesssim 5L^\ast$ galaxies. 
In the sub-mm, \citet{Trakhtenbrot2017} observed six $z\simeq4.8$ quasars with ALMA and found nearby companions around three of the quasars, at distances of 14--45 kpc. 
Given that these companions are not detected in the infrared with Spitzer, \citet{Trakhtenbrot2017} conclude that there must be significant dust-attenuation in these galaxies.
Continuing this study, \citet{Nguyen2020} observed an additional 12 quasars, finding nearby companions around five of the 18 quasars.
\citet{Decarli2017} detected a companion galaxy around four of 25 $z\gtrsim6$ quasars with ALMA, and similarly \citet{Willott2017} found one quasar companion in a sample of five $z\gtrsim6$ quasars. \citet{Mazzucchelli2019} took follow-up observations of these companions in the optical/IR, detecting the emission from only one of the companions, finding that the remaining three must be ``highly dust-enshrouded''. \citet{willott_2005} also hypothesise that the lack of companions observed in rest-frame UV observations, relative to sub-mm observations, is a result of dust attenuation.

In Figure \ref{NumNeighbours} we consider the effect of dust attenuation on the observed number of companion galaxies, by showing the number of companions that are intrinsically brighter than each magnitude limit, and the fraction of these companions that have dust-attenuated magnitudes fainter than the limit and thus would be `missed' by observations in the rest-frame UV.
We find that at a depth of $m_{\textrm{UV}}=22.8$, almost 100 per cent of companions with intrinsic magnitudes of $m_{\textrm{UV}}<22.8$ are missed due to dust attenuation (with $m_{\textrm{UV, Dust}}>22.8$). However, the overall number of these intrinsic companions is low (an average of $<1$ within 300 kpc).
At a magnitude limit of $m_{\textrm{UV}}=24.85$, around 50 per cent of the companions of the most massive black holes are missed, while for the quasar samples this is around 75 per cent. 
RST, at a depth of $m_{\textrm{UV}}=26.9$, will be able to detect the majority of companions, with less than 10 per cent of intrinsic companions missed due to dust attenuation.

Overall, our predictions expect that a large fraction (up to 75\% at $m_{\textrm{UV}}<24.85$) of quasar companions will be `missed' in current rest-frame UV observations due to dust obscuration. These dusty galaxies are likely to be observable in the sub-mm, and so our predictions are consistent with expectations \citep[e.g][]{willott_2005}.

\subsection{Properties of nearby neighbours}

\begin{figure*}
\includegraphics[scale=0.91]{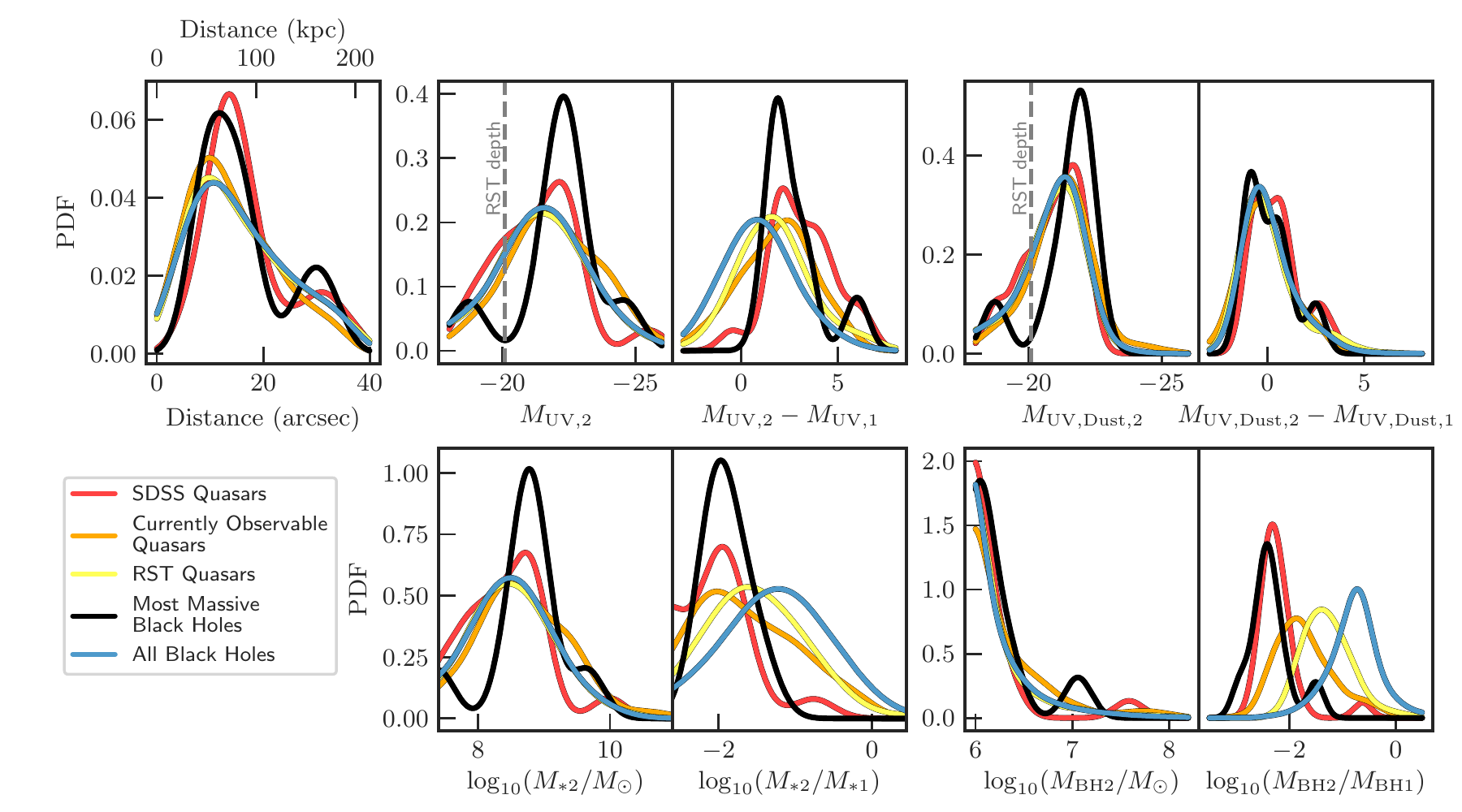}
\begin{center}
\caption{The properties of the nearest neighbour to each black hole in the various samples, that have a distance of less than 200 kpc. The various panels show probability distribution functions for the distance to the nearest neighbour in arcseconds and kpc, the UV magnitude of the neighbouring galaxy (with and without dust-attenuation), the difference between the host and companion's magnitude, the companion's stellar and black hole mass, and the stellar and black hole mass ratio of the neighbour's mass to that of the sample black hole's host. The `all black hole' sample refers to black holes with $M_{\textrm{BH}}>10^{6.5}M_\odot$. These probability distribution functions are calculated using kernel density estimation with Gaussian kernels, and a covariance factor of 0.4 for appropriate smoothing.}
\label{Neighbours}
\end{center}
\end{figure*}

We now restrict our investigation to the nearest neighbour to each black hole, with distances less than 200 kpc.

We find that 90 per cent of the most massive black holes have their nearest neighbour  within 200 kpc, compared with 87 per cent of SDSS quasars, 80 per cent of currently observable quasars and 67 per cent of RST quasars. For comparison, 63 per cent of all black holes with $M_{\textrm{BH}}>10^{6.5}M_\odot$ have their nearest neighbour within 200 kpc.

Figure \ref{Neighbours} shows various properties of the nearest neighbours: their distance, UV magnitude (both with and without dust attenuation), stellar mass and black hole mass, and the differences between the properties of the neighbour and those of the black hole host.
Most of the nearest neighbours lie within 100 kpc or 20 arcseconds of the black hole host galaxy.
The vast majority of these neighbours are brighter than $M_{\textrm{UV}}=-20$, so should be readily detectable by RST. Some companions are fainter than the black hole host by up to 5 magnitudes, although most are of similar brightness.
The stellar mass and black hole mass distributions of the neighbouring galaxies are consistent between the various black hole samples. However, as the most massive black holes and quasars are hosted by massive galaxies, the stellar mass ratios between the neighbour and the black hole host $M_{\ast2}/M_{\ast1}$ are lower. 
More than 75 per cent of the neighbours of quasars, and 90 per cent of the neighbours of the most massive black holes, have less than 1/10th of the stellar mass of the black hole host, and so would be classified as only minor mergers; this is compared with around 60 per cent of the neighbours of all $M_{\textrm{BH}}>10^{6.5}M_\odot$ black holes.
More than 75 per cent of the neighbours of quasars and most massive black holes have black hole mass ratios $M_{\textrm{BH}2}/M_{\textrm{BH}1}$ that are also less than 1/10, compared with 26 per cent of neighbours of all $M_{\textrm{BH}}>10^{6.5}M_\odot$ black holes.

The quasar companions observed by \citet{Trakhtenbrot2017} have dynamical masses $M_{\textrm{dyn}}=(2.1-10.7) \times 10^{10} M_\ast$, relative to $M_{\textrm{dyn}}=(3.7-7.4) \times 10^{10} M_\ast$ for the quasar hosts, and so these interactions would be classified as major mergers.
The quasar companions found by \citet{McGreer2014} are also likely to be major mergers. These companions are also found at projected distances of 5 and 12 kpc, which, depending on the angle of projection, are much smaller than the average distance of companions in the \BlueTides simulation, as are the majority of ALMA-discovered companions, due to its small field of view. This may be a result of the companion classification used in the simulation being ineffective at low separations.

\begin{figure*}
\includegraphics[scale=0.43]{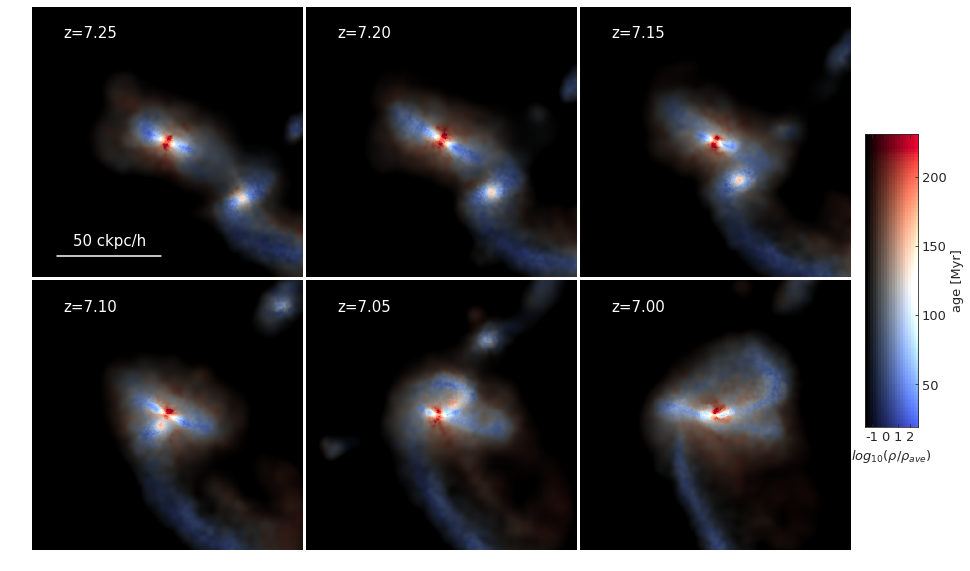}
\begin{center}
\caption{The host galaxies of two merging black holes from $z=7.25$ to $z=7.0$. 
At $z=7.3$, the primary black hole (at the centre of the images) has a black hole mass of $\log(M_{\textrm{BH}}/M_\odot)=8.25$, while the companion has a mass of $\log(M_{\textrm{BH}}/M_\odot)=6.37$ and is at a distance of 82 ckpc or 9.9 kpc from the primary black hole.
The black hole which results from this merger is one of the ten most massive black holes in the simulation at $z=7$, with a mass of $\log(M_{\textrm{BH}}/M_\odot)=8.56$, in a galaxy with a stellar mass of $\log(M_\ast/M_\odot)=11.11$. 
Each panel is 120$/h$ ckpc per side, showing the stellar density colour coded by the age of star (from blue to red indicating young to old populations respectively).}
\label{MergerImages}
\end{center}
\end{figure*}

We examine one of the most massive black holes which has its nearest neighbour within 200 kpc more closely.
This black hole has a mass of $\log(M_{\textrm{BH}}/M_\odot)=8.56$, and is hosted by a galaxy of mass $\log(M_{\ast}/M_\odot)=11.11$.
We find that at $z=7.0$, its dark matter halo contains 5 additional black holes, with masses $\log(M_{\textrm{BH}}/M_\odot)=(8.20,7.80, 6.53,7.06,5.97)$, in galaxies with stellar masses of $\log(M_\ast/M_\odot)=(10.54,10.58,9.24,9.66,8.57)$.
%10.76,10.83,9.51,9.93,8.89). - Full 'orig' masses
Imaging this system at various redshifts shows that this galaxy has been involved in a recent merger between $z=7.3$ and $z=7.0$, with the central black hole (of mass $\log(M_{\textrm{BH}}/M_\odot)=8.25$ at $z=7.3$) merging with another black hole of mass $\log(M_{\textrm{BH}}/M_\odot)=6.37$ (Figure \ref{MergerImages}).

\section{Conclusions}
\label{sec:Conclusions}
%%%%%%%%%%%%%%%%%%%% REFERENCES %%%%%%%%%%%%%%%%%%

In this paper we use the \BlueTides simulation to make predictions for the host galaxies of the most massive black holes and quasars at $z=7$. Our main findings are as follows.
\begin{itemize}
    \item The 10 most massive black holes are in massive galaxies with stellar masses $\log(M_\ast/M_\odot)=10.80\substack{+0.20 \\ -0.16}$, which have large star formation rates, $513\substack{+1225 \\ -351}M_\odot/\rm{yr}$. Quasar hosts are less massive,  $\log(M_\ast/M_\odot)=10.25\substack{+0.40 \\ -0.37}$,
    with lower star formation rates, $191\substack{+288 \\ -120}M_\odot/\rm{yr}$. 
    Lower luminosity quasars are hosted by less extreme host galaxies.
    \item The hosts of the most massive black holes and quasars in \BlueTides are generally bulge-dominated, with $B/T\simeq0.85\pm0.1$, however their morphologies are not biased relative to the overall $z=7$ galaxy sample.
    \item The hosts of the most massive black holes and quasars are compact, with half-mass radii of $R_{0.5}=0.41\substack{+0.18 \\ -0.14}$ and $0.40\substack{+0.11 \\ -0.09}$ kpc respectively. Galaxies of similar mass and luminosity have a wider range of sizes with a larger median value, $R_{0.5}=0.71\substack{+0.28 \\ -0.25}$ kpc.
    \item Despite its increased resolution over HST, distinguishing the compact host galaxies from the quasar emission will still be challenging with JWST, as shown through our mock images. This will be more successful for galaxies that have the lowest contrast ratio between the host and the AGN.
    \item The $z=7$ sample has a black hole--stellar mass relation that is steeper than the local \citet{Kormendy2013} relation, but the two are reasonably consistent, particularly at the highest masses where the observations are most robust. Sub-mm observations of $5\lesssim z\lesssim7$ quasars with $M_{\textrm{BH}}<10^{8.5}M_\odot$ are consistent with our predicted relation.
    \item Observations of quasars are biased to measure a higher black hole--stellar mass relation than the intrinsic relation. The SDSS, currently observable, and RST quasar samples have black hole--stellar mass relations 0.2 dex higher than the total galaxy sample, providing an estimate of the systematic offset of quasar observations of the $M_\ast$--$M_{\textrm{BH}}$ relation from the true population. To reduce this bias in the measured black hole--stellar mass relation, observations should target AGN that are faint relative to their host galaxies, which are more likely to be found in deep surveys.
    \item The most massive black holes and quasars have more nearby companions than the typical $M_{\textrm{BH}}>10^{6.5}M_\odot$ black hole. The majority of their nearest neighbours have stellar mass ratios $M_{\ast2}/M_{\ast1}<0.1$ and thus would be classified as minor mergers. 
    A large fraction of these nearby companion galaxies will be missed by rest-frame UV observations due to dust attenuation.
\end{itemize}

\section*{Acknowledgements}
We thank the referee for their valuable and constructive comments which helped to improve the quality of this paper.
We also thank Bram Venemans for a useful discussion on the measurement of star formation rates from sub-mm observations.
This research was supported by the Australian Research Council Centre of Excellence for All Sky Astrophysics in 3 Dimensions (ASTRO 3D), through project number CE170100013.
The \BlueTides simulation was run on the BlueWaters facility at the National Center for Supercomputing Applications.
Part of this work was performed on the OzSTAR national facility at Swinburne University of Technology, which is funded by Swinburne University of Technology and the National Collaborative Research Infrastructure Strategy (NCRIS).
MAM acknowledges the support of an Australian Government Research Training Program (RTP) Scholarship, and a Postgraduate Writing-Up Award sponsored by the Albert Shimmins Fund.
TDM acknowledges funding from NSF ACI-1614853, NSF AST-1517593, NSF AST-1616168 and NASA ATP 19-ATP19-0084.
TDM and RAC also acknowledge ATP 80NSSC18K101 and NASA ATP 17-0123.
This paper makes use of version 17.00 of Cloudy, last described by \citet{Ferland2017}.

\section*{Data Availability}
The data underlying this article will be shared on reasonable request to the corresponding author.

\bibliographystyle{mnras}
\bibliography{Paper.bib} 

%%%%%%%%%%%%%%%%%%%%%%%%%%%%%%%%%%%%%%%%%%%%%%%%%%

% Don't change these lines
\bsp	% typesetting comment
\label{lastpage}
\end{document}